\documentclass{aa}     

\usepackage{graphicx}
\usepackage{txfonts}
\usepackage{natbib}
\usepackage{url}
\bibpunct{(}{)}{;}{a}{}{,}

\newcommand{\be}{\begin{equation}}
\newcommand{\ee}{\end{equation}}

\def\apjl{ApJL}
\def\apjs{ApJS}

\newcommand{\Mpc}{$h^{-1}$\thinspace Mpc}

\begin{document}

\title{Multimodality in galaxy clusters from SDSS DR8:
substructure and velocity distribution
}

\author {M.~Einasto\inst{1} \and  J.~Vennik\inst{1}  \and P.~Nurmi\inst{2} 
 \and E.~Tempel\inst{1,3} \and A.~Ahvensalmi\inst{2} \and E.~Tago\inst{1}
 \and L.J.~Liivam\"agi\inst{1,4} \and E.~Saar\inst{1,5}  \and  P.~Hein\"am\"aki\inst{2} 
\and J.~Einasto\inst{1,5,6} \and V.J.~Mart\'{\i}nez\inst{7}
}

\institute{Tartu Observatory, 61602 T\~oravere, Estonia
\and 
Tuorla Observatory, University of Turku, V\"ais\"al\"antie 20, Piikki\"o, Finland
\and 
National Institute of Chemical Physics and Biophysics, Tallinn 10143, Estonia
\and
Institute of Physics, Tartu University, T\"ahe 4, 51010 Tartu, Estonia
\and
Estonian Academy of Sciences,  EE-10130 Tallinn, Estonia
\and 
ICRANet, Piazza della Repubblica 10, 65122 Pescara, Italy
\and 
Observatori Astron\`omic, Universitat de Val\`encia, Apartat
de Correus 22085, E-46071 Val\`encia, Spain
}

\authorrunning{M. Einasto et al. }

\offprints{M. Einasto}

\date{ Received   / Accepted   }

\titlerunning{Substructure of clusters}

\abstract
{
The study of the signatures
of multimodality in groups and clusters of 
galaxies, an environment for most of the galaxies in the Universe,
gives us information about the dynamical state of clusters and about merging 
processes, which affect the formation and evolution
of galaxies, groups and clusters, and larger structures -- superclusters
of galaxies and the whole cosmic web.
}
{
We search for the presence of substructure, 
a non-Gaussian, asymmetrical velocity distribution of galaxies, 
and large peculiar velocities of the main galaxies
in clusters with at least 50 member
galaxies, drawn from the SDSS DR8.
}
{
We employ a number of 3D, 2D, and 1D tests to analyse the distribution of 
galaxies in clusters: 3D normal mixture modelling,
the Dressler-Shectman  test, the Anderson-Darling and 
Shapiro-Wilk  tests, as well as the Anscombe-Glynn and the D'Agostino tests.  We 
find the peculiar velocities of the main galaxies, and use principal 
component analysis to characterise our results. 
}
{
More than 80\% of the clusters in our sample have substructure
according to 3D normal mixture modelling, and
the Dressler-Shectman (DS) test shows 
substructure in about 70\% of the clusters.  
The median value of the 
peculiar velocities of the main galaxies in clusters is 206 km/s (41\% of the rms velocity).
The velocities of galaxies in more than 20\% of the clusters
show significant non-Gaussianity. 
While multidimensional normal mixture modelling 
is more sensitive than the DS test in resolving substructure
in the sky distribution of cluster galaxies, the DS test determines better
substructure expressed as tails in the velocity distribution of galaxies
(possible line-of-sight mergers).
Richer, larger, and more luminous clusters have larger amount of substructure and 
larger (compared to the rms velocity) peculiar velocities of the main galaxies. 
Principal component analysis of both the substructure indicators and the physical 
parameters of clusters shows that galaxy clusters are complicated 
objects, the properties of which cannot be explained with a small number of parameters
or delimited by one single test. 
}
{
The presence of substructure, the non-Gaussian velocity distributions, 
as well as the large peculiar velocities of the main 
galaxies, shows that most of the clusters in our sample are dynamically young. 
}
%\end{abstract}

\keywords{Cosmology: large-scale structure of the Universe;
galaxies: clusters: general}

\maketitle

\section{Introduction} 
\label{sect:intro} 

Most galaxies in the Universe are located in groups and clusters of galaxies, 
which themselves reside in larger systems -- in superclusters of galaxies or 
in filaments crossing underdense regions between superclusters 
\citep{1978MNRAS.185..357J, 1978ApJ...222..784G, zes82, 1986ApJ...302L...1D}. In 
the $\Lambda$CDM concordance cosmological model the structures forming the 
cosmic web grow by hierarchical clustering driven by gravity \citep[see, e.g.,][and 
references therein]{2002PhRvD..65d7301L, loeb2008}. The present-day dynamical 
state of clusters of galaxies depends on their formation history. Signatures of 
multimodality in the distribution of galaxies in clusters (the presence of 
substructure, several galaxy associations within clusters, non-Gaussian 
velocity distributions of galaxies, and large peculiar velocities of the main 
galaxies) are indicators of former or ongoing mergers in 
groups and clusters, which affect the formation and evolution of galaxies 
\citep{1993AJ....105.1596B, 
1996ApJS..104....1P, 2000A&A...354..761K}. These mergers 
have shaped the properties of galaxies in groups 
and clusters, as, e.g., the well-known morphological segregation effect
\citep{1974Natur.252..111E,1980ApJ...236..351D, 1987MNRAS.226..543E, 
2009ApJ...690.1292B, 2009A&A...505...83H}. Substructure affects 
estimates of several cluster characteristics, the dynamical mass and 
mass-to-light ratio among others 
\citep[][and references therein]{2006A&A...456...23B, niemi, 
2008A&A...491...71P, 2008MNRAS.383..720H,
2010MNRAS.408.1818W, 2011MNRAS.tmp.1734P}. 
Detailed knowledge of the properties of clusters of galaxies 
is needed for comparison of observations with N-body models of the 
formation and evolution of cosmic structures, to test the cosmological 
models \citep{1998MNRAS.296.1061T, 2009MNRAS.400.1317A}. 

To search for signatures of multimodality in galaxy clusters a number of 3D, 2D, and 1D
methods have been proposed, including the Dressler-Shectman test 
\citep{dressler88,2000A&A...354..761K}, the hierachical clustering method
\citep{1996A&A...309...65S, 2010A&A...517A..94D}, wavelet analysis \citep{2006A&A...450....9F},
multidimensional normal mixture modelling \citep{fraley2006}, 
see also \citet{2010A&A...522A..92E}, 
and many others \citep[see ][for a review]{1996ApJS..104....1P}.
\citet{1996ApJS..104....1P} showed that it is preferable to use several
methods to search for substructure in clusters, since their sensitivity to 
different signatures of multimodality is different.

Several studies have shown  the presence of 
substructure in poor and rich groups and clusters of galaxies
\citep{1999A&A...343..733S, oegerlehill2001, 2001MNRAS.320...49K, bur04, 
2006A&A...449..461B, 2006A&A...450....9F,
2007A&A...469..861B, 2007ApJ...662..236H,
2008A&A...487...33B, 2009ApJ...702.1199H, 2009AN....330..998V, 2010A&A...521A..28A,
2011MNRAS.410.1837P, 2011MNRAS.413L..81R, 2012MNRAS.419L..24M, 2012arXiv1201.3676H}, 
and in X-ray clusters \citep{2007A&A...470...39R, 2009ApJ...692..702O, 
2009ApJ...693..901O, 2010A&A...514A..32B}. 
\citet{tov09} studied the properties  of poor groups from the SDSS survey and 
showed that many groups of galaxies are not presently in a dynamical equilibrium, but 
at various stages of virialization. \citet{niemi} showed that a significant 
fraction of nearby groups of galaxies are not even gravitationally bound systems.

In this paper we  study  the multimodality in rich clusters drawn from the SDSS DR8.
We use data about 109 clusters with at least 50 
member galaxies, this is one of the largest samples of
rich clusters analysed for substructure so far. 
We employ a number of 3D, 2D, and 1D methods to analyse the distribution of 
galaxies in clusters. With the principal component analysis we characterise 
the results of different tests simultaneously, and study the relations
between the multimodality of clusters and their physical properties. We   
present lists of unimodal and multimodal clusters.
In Sect.~\ref{sect:data} we describe the data we used. 
In Sect.~\ref{sect:methods} we describe the methods to search for signatures of
multimodality, and we apply them in Sect.~\ref{sect:results} 
to study the properties of clusters. We discuss the results 
and draw conclusions in Sect.~\ref{sect:discussion}.

We assume  the standard cosmological parameters: the Hubble parameter $H_0=100~ 
h$ km s$^{-1}$ Mpc$^{-1}$, the matter density $\Omega_{\rm m} = 0.27$, and the 
dark energy density $\Omega_{\Lambda} = 0.73$.

\section{Data} 
\label{sect:data}

We used the MAIN galaxy sample of the 8th data release of the Sloan
Digital Sky Survey \citep{2011ApJS..193...29A}
with the apparent $r$ magnitudes $r \leq 17.77$, 
and the redshifts $0.009 \leq z \leq 0.200$, in total 576493 galaxies.  
We corrected the redshifts
of galaxies for the motion relative to the CMB and computed the co-moving
distances \citep{mar03} of galaxies.
The absolute magnitudes of galaxies were determined in the $r$-band ($M_r$) with the
$k$-corrections for the SDSS galaxies, calculated using the KCORRECT
algorithm \citep{blanton03a,2007AJ....133..734B}. In addition, we
applied evolution corrections, using the luminosity evolution model of
\citet{blanton03b}. The magnitudes correspond to the rest-frame at the
redshift $z=0$. More details on similar data reduction for the SDSS DR7 can be found in 
\citet[][hereafter T10]{2010A&A...514A.102T}.

We determine groups of galaxies using the Friends-of-Friends (FoF)
cluster analysis 
method introduced in cosmology by \citet{tg76,zes82,hg82},  
and modified in \citet{tago08} and in T10. 
A galaxy belongs to a group of galaxies if this galaxy has at least one group 
member galaxy closer than a linking length. In a flux-limited sample the density 
of galaxies slowly decreases with distance. To take this selection effect into 
account properly when constructing a group catalogue from a flux-limited sample, 
we rescaled the linking length  with distance, calibrating the scaling relation 
by observed groups (see T10 for details). As a result, the maximum sizes in the 
sky projection and the velocity dispersions of our groups are similar at all 
distances. This shows that distance-dependent selection effects have been 
properly accounted for. Our catalogue contains 77858 groups with at least 2 
member galaxies. The richest groups in our catalogue correspond to rich clusters 
of galaxies.
The details and availablility 
of the group catalogue based on SDSS DR8 are described in \citet{2011arXiv1112.4648T}.

In flux-limited samples galaxies outside
the observational window remain unobserved. To calculate the total
luminosities of groups we have  to take into account
the luminosities of these galaxies as well.  For that, we multiply the observed galaxy
luminosities by the luminosity weight $W_d$.  The distance-dependent weight
factor $W_d$ was calculated as follows:
\begin{equation}
    W_d =  {\frac{\int_0^\infty L\,n
    (L)\mathrm{d}L}{\int_{L_1}^{L_2} L\,n(L)\mathrm{d}L}},
    \label{eq:weight}
\end{equation}
where $L_{1,2}=L_{\sun} 10^{0.4(M_{\sun}-M_{1,2})}$ are the luminosity 
limits of the observational window at a distance $d$, corresponding to the 
absolute magnitude limits of the survey $M_1$ and $M_2$; we took 
$M_{\sun}=4.64$\,mag in the $r$-band \citep{2007AJ....133..734B},
and n(L) is  the galaxy luminosity function. 
Owing to their peculiar velocities, 
the distances of galaxies are somewhat uncertain; if the galaxy belongs to a 
group, we used the group distance (mean distance of galaxies in a group)
to determine the weight factor. 
Detailed description, how the weight factor and luminosity function are
calculated can be found in \citet{2011A&A...529A..53T}.

In the group catalogue the main galaxy of a group is defined
as the most luminous galaxy in the $r$-band. We use this definition also
in the present paper. 

\begin{figure}[ht]
\centering
\resizebox{0.45\textwidth}{!}{\includegraphics[angle=0]{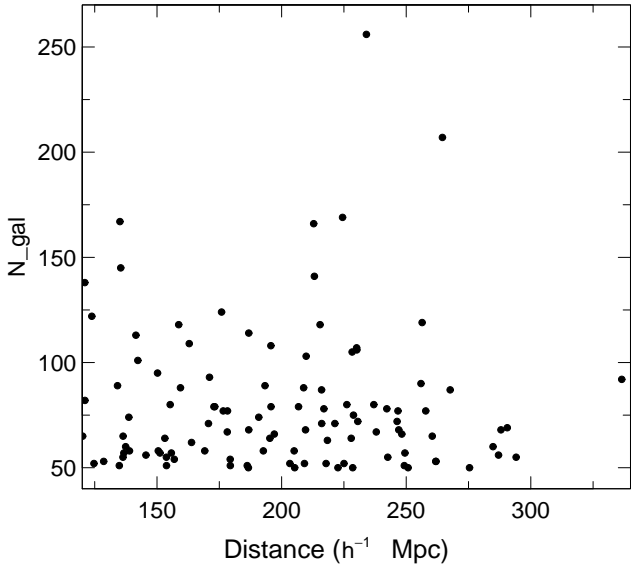}}
\caption{
Richness of clusters vs. their distance.  
}
\label{fig:richdist}
\end{figure}

Next we select clusters for this study. The larger the number of galaxies 
in clusters, the more reliable is the analysis of substructure
and their velocity distribution. \citet{2010A&A...521A..28A} select clusters with 
at least 30 member galaxies for substructure study. However, 
\citet{2008A&A...487...33B} argue that 
clusters with about 30 member galaxies are too small for the analysis 
of substructure \citep[see also the discussion about the substructure statistics 
in small samples in][]{2006A&A...456...23B}, although 
clusters with smaller numbers of galaxies have been studied
for substructure \citep[see, for example,][]{1999A&A...343..733S,
2008A&A...487...33B, 2011MNRAS.413L..81R}. 
Another problem arises with the selection effects: in the  group catalogue 
the richness of groups decreases rapidly at distances $D > 340$~\Mpc\ owing to 
the use of a flux-limited sample of galaxies (T10). At distances smaller than 
120~\Mpc\ the sample includes  nearby exceptionally rich clusters which 
correspond to well-known Abell clusters \citep[the Coma cluster, rich clusters in the 
Hercules supercluster and others, see T10 and][]{2011A&A...532A...5E}. These 
clusters have to be analysed separately. Therefore  we chose for the present 
analysis clusters with at least 50 member galaxies in the distance interval 
120~\Mpc\ $\le D \le $ 340~\Mpc\
(redshift interval $0.04 \leq z \leq 0.12$). 
This sample includes all clusters from the SDSS DR8 
with at least 50 member galaxies from our catalogue in this distance interval, 
in total 109 clusters. 
Figure~\ref{fig:richdist} shows the richness of clusters in our sample vs. their 
distance.

\begin{figure}[ht]
\centering
\resizebox{0.45\textwidth}{!}{\includegraphics[angle=0]{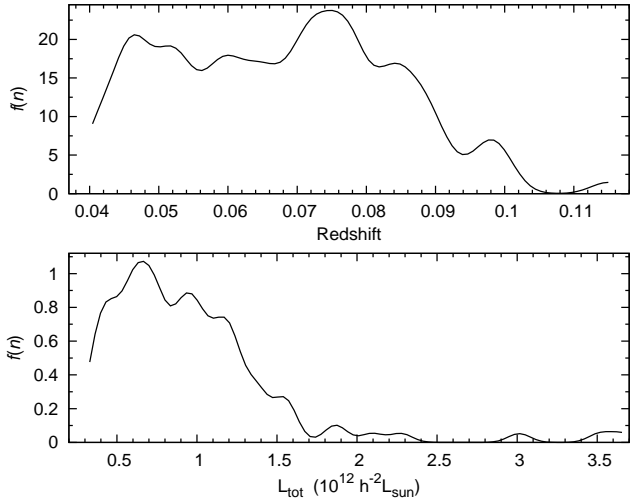}}
\caption{
Distribution of cluster redshifts (upper panel) and 
total luminosities (lower panel).  
}
\label{fig:redlum}
\end{figure}

We cross-identify groups with Abell clusters, which have acquired a role of a 
reference system for rich clusters. Contrary to expectations, cross-
identification of (rich) SDSS DR8 groups with Abell clusters is not 
straightforward. Problems arise both due to different group/cluster finding 
procedures as well as different algorithms for components/subgroups.  The Abell 
clusters have a constant  linear radius (1.5~\Mpc), while the DR8 groups 
obtained by a FoF procedure have various linear sizes. An Abell cluster may 
consist of several subclusters and/or may be the result of projections of groups 
from different  distance \citep[e.g.][] {2011MNRAS.410.1837P}. These  facts make 
cross-identification  difficult.  We consider a group identified with an 
Abell cluster, if the distance between their centres is smaller than at least 
the linear radius of one of the clusters, and the distance between their centres 
in the radial (line-of-sight) direction is less than 600 km/s (an empirical 
value). As a result one group can be identified with more than one Abell 
clusters and vice versa. This can be seen  in  online tables, where we give data 
on the clusters, and the results of the tests
(Table~\ref{tab:cldata1} and Table~\ref{tab:testresults1}). 
In Fig.~\ref{fig:redlum} we show the distributions of cluster redshifts
and total luminosities, as given in Table~\ref{tab:cldata1}.

\section{Methods}
\label{sect:methods} 

In this section we decribe the methods applied in this paper to analyse the 
multimodality of galaxy clusters.

\subsection{Multidimensional normal mixture modelling with {\it Mclust}}
\label{sect:mclust}

To search for possible components in clusters, we employ multidimensional normal 
mixture modelling based on the analysis of a finite mixture of distributions, in 
which each mixture component is taken to correspond to a different group, 
cluster or subpopulation. The most common component distribution considered in 
model-based clustering is a multivariate Gaussian (or normal) distribution.  To 
model the collection of components, we apply the {\it 
Mclust} package for classification and clustering
\citep{fraley2006} from {\it R}, an open-source free statistical environment 
developed under the GNU GPL \citep[][\texttt{http://www.r-project.org}]{ig96}. 
This package searches for an 
optimal model for the clustering of the data among models with varying shape, 
orientation and volume, finds the optimal number of components, and the 
corresponding classification (the membership of each component).
{\it Mclust} calculates for every galaxy the probabilities to belong to any of 
the components. The uncertainty of classification is defined as one minus the 
highest probability of a galaxy to belong to a component. The mean uncertainty 
for the full sample is used  as a statistical estimate of the reliability of the 
results.

We tested how the possible errors in the line-of-sight positions of galaxies 
affect the results  of {\it Mclust}, shifting randomly the peculiar velocities 
of galaxies 1000 times and searching  each time for the components with {\it 
Mclust}. The random shifts were chosen from a Gaussian
distribution with the dispersion equal to the sample velocity dispersion of galaxies 
in a cluster.  The number of the components found by {\it Mclust} remained unchanged, 
demonstrating that the results of {\it Mclust} are not sensitive to such errors.

We also performed substructure analysis with {\it Mclust}
using only the sky coordinates of the clusters as an additional 2D test.

\subsection{DS test}

Another diagnostic for substructure in clusters is 
the Dressler-Shectman (DS or $\Delta$) test 
\citep{dressler88}. The DS test 
searches for deviations of the 
local velocity mean and dispersion from the cluster mean values. The algorithm 
starts by calculating the mean velocity ($v_\mathrm{local}$) and the velocity dispersion 
($\sigma_\mathrm{local}$) for each galaxy of the cluster, using its $n$ nearest 
neighbours. These values of local kinematics are compared with the mean velocity 
($v_c$) and the velocity dispersion ($\sigma_c$) determined for the entire cluster of $N_{gal}$ 
galaxies. The differences between the local and global kinematics are 
quantified by 

$$ \delta_i^2 = (n+1)/\sigma_c^2\left[(v_\mathrm{local}-v_c)^2 + (\sigma_\mathrm{local}-\sigma_c)^2\right]. $$

The cumulative deviation $\Delta = \Sigma \delta_i$ is  used as a statistic 
for quantifying (the significance of) the substructure.
The results of  the DS-test depend on the number of local galaxies $n$. Here, we have studied the
substructure by using $n = \sqrt N_{gal}$, as suggested by \citet{1996ApJS..104....1P}.

If the cluster velocity distribution is close to a Gaussian, then  $\Delta$ will be 
of order $N_{gal}$. For the case of non-Gaussian velocities, $\Delta$ can differ significantly from 
$N_{gal}$, even if there is no subclustering \citep{dressler88}. Therefore, the 
$\Delta$ statistic for each cluster should be calibrated by Monte Carlo 
simulations. In Monte Carlo models the velocities of galaxies are randomly 
shuffled among the positions, which effectively destroys any true correlation 
between the velocities and positions. We ran 25000 models for each cluster
and calculated every time $\Delta_\mathrm{sim}$. The 
significance of having substructure (the $p$-value) can be quantified by the ratio 
$N(\Delta_\mathrm{sim} > \Delta_{obs})/N_\mathrm{sim}$ --  
the ratio of the number of simulations in which the value of $\Delta$ is larger than 
the observed value, and the total number of simulations. 
The smaller the $p$-value, the larger is the 
probability of substructure.

\subsection{$\alpha$ test}

With the $\alpha$ test developed by \citet{1990ApJ...350...36W} we analyse 
correlations between positions and velocities of galaxies and
search for a region which differs from the overall distribution. This test gives
a measure of the centroid shift for the cluster galaxies.
The idea of the test 
is to search for a region that shows correlation between the positions and
velocities that differs from that for the overall galaxy distribution.
The method consists of a five different steps.
At first the centroid of the system $(x_c,y_c)$ is calculated as a
global mean value. Next each galaxy $i$ gets a weight $w_i=1/V_i$,
where $V_i$ is the line-of-sight rms velocity  calculated
using the velocities of the galaxy and its 10 nearest members.
For each galaxy 10 nearest neighbours are chosen in the velocity
space and new centroid $(x_c^{'},y_c^{'})$ is calculated using the weights:
\begin{equation}
	x_c^{'} = \frac{\sum_{i=1}^{11}x_i w_i}{\sum_{i=1}^{11}w_i},\quad y_c^{'} = \frac{\sum_{i=1}^{11}y_i w_i}{\sum_{i=1}^{11}w_i}.
\end{equation}
In the fourth step it is calculated how much this new centroid differs
from the global centroid:
\begin{equation}
	\alpha_i=\sqrt{(x_c-x_c^{'})^2+(y_c-y_c^{'})^2}.
\end{equation}
Finally the average value of $\alpha_i$ for all galaxies is defined as
the $\alpha$-value.
This is a measure of how much the centroid of all galaxies
shifts as a result of local correlations between the positions and velocities
of galaxies.

\subsection{$\beta$ test}

We also employ 2D tests which use information about the sky positions
of galaxies in clusters: the 
$\beta$ test which studies the asymmetry in the galaxy
distribution \citep{1988ApJ...327....1W} and  2D normal mixture 
modelling with {\it Mclust}.

The $\beta$ test, presented in \citet{1988ApJ...327....1W}, 
is a test for the asymmetry in the galaxy
distribution.
At first, the mean distance $d_i$ for each galaxy $i$ to its
five nearest neighbours is calculated.  Then a point diametrically opposite to this
point is chosen and the mean distance $d_o$ in this neighbourhood is
calculated as before. The asymmetry is then measured by the $\beta$-value:
\begin{equation}
	\beta = \log_{10}(d_o/d_i).
\end{equation}
To evaluate the asymmetry the average value $<\beta>$ over all galaxies is
calculated and any deviation from  $<\beta>\approx 0$ signifies
asymmetry and possible substructure.
According to \citet{1996ApJS..104....1P} the $\beta$ test is sensitive to the 
mirror asymmetry, but not to the deviations from the radial symmetry.
The $p$-values for the $\alpha$ and $\beta$ tests are calculated 
with Monte Carlo simulations 
as for the DS test using 25000 random sampling of galaxy coordinates in clusters.

\subsection{1D tests}

One indicator of multimodality in clusters is the deviation of the distribution 
of the galaxy velocities in clusters  from a Gaussian. 
To analyse the distribution of the velocities of galaxies 
we use several 1D tests. We tested the hypothesis about the Gaussian 
distribution of the peculiar velocities of galaxies in clusters with the 
Shapiro-Wilk normality test \citep{shapiro65}, which is 
considered the best for small samples. We also employed the Anderson-Darling
test, which is very reliable according to \citet{2009ApJ...702.1199H}.
We calculated the kurtosis and the 
skewness of the peculiar velocity distributions, and used these to test for the 
asymmetry of the distributions of galaxy velocities.
We used the Anscombe-Glynn test for the kurtosis 
\citep{anscombe83}, and  the D'Agostino test for the skewness 
\citep{agostino70}, from the $R$ package {\it moments} by L. Komsta and F. 
Novomestky.

There is, of course, no reason to assume that the velocity distributions in 
galaxy clusters should be exactly Gaussian. In fact, kinematic models show that 
the shape of the velocity distribution is defined by the ratio of different 
types of galaxy orbits \citep{1987ApJ...313..121M}. The tests described above 
serve mostly to check if the velocity distribution is unimodal and symmetrical. 
Historically, they have been selected because of the easy availability of 
statistical tests for the Gaussian distribution; we use them to be able to 
compare our results with those obtained earlier.

In virialized clusters galaxies follow the cluster potential well. If so, we 
would expect that the main galaxies in clusters lie at the centres of 
groups (group haloes) and have small peculiar velocities 
\citep{ost75,merritt84,malumuth92}. Therefore the peculiar velocity  of the 
main galaxies in clusters is also an indication of the 
dynamical state of the cluster \citep{coziol09}.  
We calculate the peculiar velocities of the main galaxies, 
$|v_{\mathrm{pec}}|$, the normalised peculiar velocities of the main 
galaxies, $v_{\mathrm{pec,r}} = |v_{\mathrm{pec}}|/{\sigma}_{v}$, and 
analyse the location of the main galaxies in the clusters and subclusters. 

\subsection{Principal component analysis}

We employ the principal component analysis (PCA) to analyse the results of all 
tests simultaneously, and to study the relations between the multimodality 
indicators and the physical parameters of clusters. The aim of the PCA is to study 
the relations between the parameters and, if possible,  to find a small number of 
linear combinations of correlated parameters  to describe most of the variation 
in the dataset with a small number of new uncorrelated parameters. The PCA 
transforms the data to a new coordinate system, where the greatest variance by 
any projection of the data lies along the first coordinate (the first principal 
component), the second greatest variance -- along the second coordinate, and so 
on. There are as many principal components as there are parameters, but often 
only the first few are needed to explain most of the total variation. 

The principal components PC$i$  ($i \in \mathbb{N}$, $i \leq N_{\mathrm{tot}}$) 
are linear combinations of the original parameters:

\begin{equation}\label{eq:pc}
 PCi = \sum_{k=1}^{N_{\mathrm{tot}}} a(k)_{i} V_{k},
\end{equation}
where $-1 \leq a(k)_i \leq 1$ are the coefficients of the linear transformation,
$V_k$ are the original parameters and $N_{\mathrm{tot}}$ is the number of the original 
parameters. In the analysis the parameters are  standardised -- they
are centred on their means, 
$ V_{k} - \overline{V_{k}}$, and normalised, divided by their standard deviations,
$\sigma( V_{k})$. 
The PCA is suitable tool for exploratory data analysis, 
to study simultaneously correlations between a large number of parameters
\citep[see][for references about applications of the PCA in astronomy]{2011A&A...535A..36E}.

\section{Results}
\label{sect:results} 

We show in Table~\ref{tab:results} the numbers and fractions of clusters with 
substructure, and with the $p$-values showing statistically significant deviations 
from Gaussianity in their galaxy distributions ($p \leq 0.05$).

\begin{table}[ht]
\caption{Results of the tests.}
\begin{tabular}{lcccccc} 
\hline\hline 
$\mathrm{Population}$                   & $N_{\mathrm{cl}}$ &\%  & $N_{\mathrm{cl}}$ &\%  & $N_{\mathrm{cl}}$ &\%\\
\hline
                               &\multicolumn{2}{c}{All}   
&\multicolumn{2}{c}{$N_{\mathrm{comp3D}} = 1$} &\multicolumn{2}{c}{$N_{\mathrm{comp3D}} > 1$} \\  
\\    
All clusters                            &    109  &    & 17 & 16 &  92  & 84  \\
\\
$N_{\mathrm{comp2D}} > 1$               &    82   & 75 &  \phantom{0}3 & 18 &  79   & 86   \\
\\                                                       
$v_{\mathrm{pec}} \geq 250$             &    48   & 44 &  \phantom{0}4 & 24 &  44   & 48  \\
$v_{\mathrm{pec,r}} \geq 0.5$           &    45   & 41 &  \phantom{0}4 & 24 &  41   & 44  \\
\\
median($v_{\mathrm{pec}}$)              &     206  &    & 149  &    &  223     &  \\
median($v_{\mathrm{pec,r}}$)            &    0.41  &    & 0.28 &    &  0.46    &   \\
\\                                            
$p_{\mathrm{\Delta}} \leq 0.05$                  &    77   & 71 &  \phantom{0}7 & 41 &    70   & 76 \\
$p_{\mathrm{\alpha}} \leq 0.05$                      &    70   & 64 &  \phantom{0}5 & 29 &    65   & 71 \\
$p_{\mathrm{\beta}} \leq 0.05$                       &    63   & 58 &  \phantom{0}1 &  \phantom{0}6 &    62   & 67 \\
$p_{\mathrm{AD}} \leq 0.05$             &    24   & 22 &  \phantom{0}3 & 18 &    21   & 23 \\
$p_{\mathrm{SW}} \leq 0.05$             &    29   & 27 &  \phantom{0}3 & 18 &    26   & 28 \\
$p_{\mathrm{kurtosis}} \leq 0.05$       &    13   & 11 &  \phantom{0}0 & \phantom{0}0  &    13   & 14 \\
$p_{\mathrm{skewness}} \leq 0.05$       &     \phantom{0}1   & \phantom{0}1  &  \phantom{0}0 & \phantom{0}0  &     \phantom{0}1   & \phantom{0}1  \\
\\
\hline            
                
\label{tab:results}  
\end{tabular}\\
\tablefoot{ 
Numbers and fractions 
of clusters in various populations and median values of the peculiar
velocities of main galaxies, $v_{\mathrm{pec}}$, (in km/s), 
and of the normalised peculiar velocities, $v_{\mathrm{pec,r}}$.
}
\end{table}

Table~\ref{tab:results} shows that more than 80\% of the clusters in our sample 
have multiple components, as detected by {\it Mclust} by the 3D analysis. About 
75\% of the clusters continue to show multiple components, if we use only the 
sky distribution of galaxies. More than 70\% of all the clusters and more than 
75\% of the multicomponent clusters show statistically significant substructure 
according to the DS test, also; these fractions are slightly smaller for the 
$\alpha$ and $\beta$ tests. The fractions of clusters with non-Gaussian velocity 
distributions are much smaller. The reason is that in many clusters galaxies 
from different components in the sky coordinates have similar (nearly Gaussian) 
distributions of velocities (we will show below a few examples of that).   
One-component clusters show signatures of substructure and non-Gaussian velocity 
distributions according to the DS, $\alpha$, and other tests as well, but as 
seen from Table~\ref{tab:results}, the fractions of such clusters are smaller 
than those for multicomponent clusters. There are three clusters for which 
$p_{\mathrm{AD}} \leq 0.05$, but $p_{\mathrm{SW}} \geq 0.05$, and eight clusters 
with $p_{\mathrm{SW}} \leq0.05$, but $p_{\mathrm{AD}} \geq 0.05$. 
\citet{2009ApJ...702.1199H} showed that the AD test is very reliable; in our 
study the SW test detected non-Gaussian velocity distributions in a larger 
number of clusters, but  there are also clusters for which the AD test found 
significant non-Gaussianity in the velocity distributions of galaxies, but the 
SW test did not. 

Comparison with other studies shows that \citet{2007A&A...470...39R} found 
substructure in 73\% of X-ray clusters which is very close to what we found 
by 2D normal mixture modelling. \citet{1999A&A...343..733S}, 
\citet{oegerlehill2001}, and
\citet{2010A&A...521A..28A} found with  the DS test 
that 30-50\% of clusters have significant substructure. 
\citet{2006A&A...450....9F} detected substructure in about 1/3 of Abell clusters 
studied by them using wavelet analysis. About 20-30\% of clusters have non-Gaussian 
velocity distributions, as shown with the AD test or other 1D tests 
\citep{1999A&A...343..733S, oegerlehill2001, 2009ApJ...702.1199H,
2011MNRAS.413L..81R, 2012MNRAS.419L..24M}.

\subsection{Peculiar velocities of cluster main galaxies}

\begin{figure}[ht]
\centering
\resizebox{0.5\textwidth}{!}{\includegraphics[angle=0]{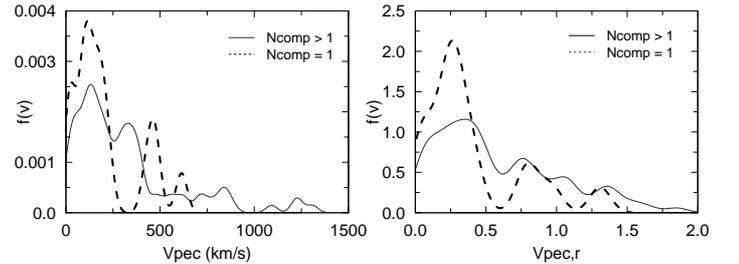}}
\caption{
Distribution of the peculiar velocities of main galaxies, $v_{\mathrm{pec}}$ (left
panel), and the normalised peculiar velocities of main galaxies, 
$v_{\mathrm{pec,r}}$ (right panel),
for clusters with one component (dashed line)
and with multiple components (solid line). 
}
\label{fig:pec}
\end{figure}

Figure~\ref{fig:pec} shows the distributions of  the peculiar velocities 
$v_{\mathrm{pec}}$
and normalised peculiar velocities $v_{\mathrm{pec,r}}$ of 
the main galaxy in clusters, separately for one-component and
multicomponent clusters. The peculiar velocity and
normalised peculiar velocity distributions for main galaxies of one-component clusters
we see a maximum at velocities less than 250 km/s (0.5 for the normalised
peculiar velocities), followed by another maximum at larger velocities. 
Checking the distributions of galaxies in  the clusters shows that in these clusters 
where the main galaxy has a peculiar velocity less than 250 km/s, 
 it is located in a component close to the 
central part of the cluster. If the peculiar velocity is higher, the main galaxy 
is located at the edge 
of the cluster, in multicomponent clusters typically in another component.  

Table~\ref{tab:results} shows that  in about half of the multicomponent clusters the 
peculiar velocity of main galaxy is larger than 250 km/s (0.5 for normalised 
peculiar velocities). There is a clear difference between the median values of the 
peculiar velocities and normalised peculiar velocities of the one-component and 
multicomponent clusters -- these velocities are much larger in the multicomponent clusters. 
This agrees with  the results of \citet{coziol09} who
found that the median value of the normalised peculiar
velocities of main galaxies in the Abell clusters is 0.32, which is between the values
found in this study.

\begin{figure}[ht] \centering 
\resizebox{0.45\textwidth}{!}{\includegraphics[angle=0]{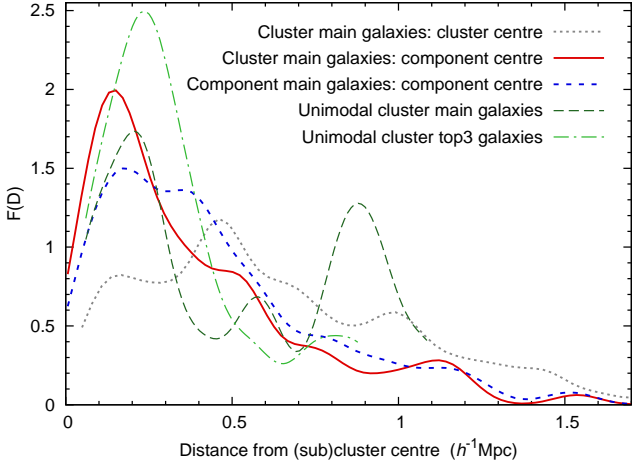}} 
\caption{ 
Distribution of the distance of the main galaxy from the cluster (subcluster or component) 
centre for various subsamples of galaxies. Multicomponent clusters:
grey dotted line -- the distance of the main galaxy from the cluster 
centre; 
red solid line -- the distance of the main galaxy from the (nearest) 
component centre; 
blue dashed line --
the distance of the brightest galaxy 
in a component from the component centre. 
One-component clusters:
dark green long-dashed line -- the distance from the cluster centre; 
light green dotted-dashed line -- the minimum distance 
of one of the three brightest galaxies from the cluster centre. 
} 
\label{fig:distmain} 
\end{figure}

Figure~\ref{fig:distmain} shows the 2D distance of the main galaxy from the cluster 
centre, both for the multicomponent and one-component clusters. It is well 
seen that in multicomponent clusters a large fraction of main galaxies are 
located far away from the cluster centre
(grey dotted line).
However, when looking at the 
components found by 3D normal mixture modelling, we see that the main galaxies 
of clusters are preferentially located close to the centre of one of the 
components
red solid line).
The distribution of distances from the component centre for the 
brightest galaxies in the components (Fig.~\ref{fig:distmain}) shows that these 
galaxies are also located preferentially close to the component centre
(blue dashed line). 

The distribution of the distances of the main galaxy from the cluster centre  in 
one-component clusters (Fig.~\ref{fig:distmain},  dark green long-dashed line) 
shows that more than a half of 
the main galaxies are located near the cluster centre, but there is also a 
substantial number of main galaxies, which are further away from the cluster 
centre. We calculated also the minimum distance from the cluster centre for 
three brightest galaxies in the one-component clusters. As seen from 
Fig.~\ref{fig:distmain} (light green dotted-dashed line), 
one of the three brightest galaxies in clusters is 
always located close to the cluster centre. This shows that the central galaxy 
of a cluster is typically one of the most luminous galaxies, but not always the 
most luminous one.

\subsection{Comparison of the results of different tests}

\begin{figure}[ht]
\centering
\resizebox{0.45\textwidth}{!}{\includegraphics[angle=0]{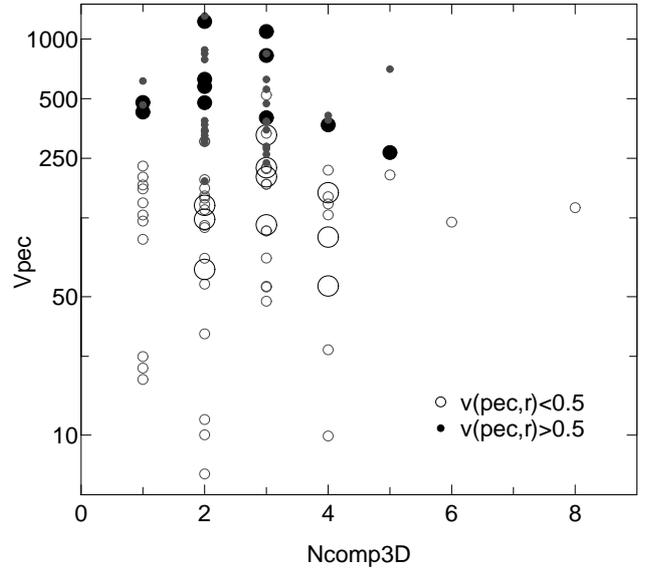}}
\caption{
Peculiar velocities of main cluster galaxies and the number of
separate components in the clusters.
Large symbols  mark clusters  with a non-Gaussian velocity distribution
of galaxies ($p_{\mathrm{AD}} < 0.05$ and $p_{\mathrm{SW}} < 0.05$). 
Empty circles denote clusters with normalised peculiar velocities of their
main galaxies $v_{\mathrm{pec,r}} < 0.5$, filled circles denote clusters
with  $v_{\mathrm{pec,r}} > 0.5$. 
}
\label{fig:ncompvpec}
\end{figure}

In Figure~\ref{fig:ncompvpec} we compare the following properties of clusters:
the number of components determined by the 3D version of {\it Mclust}, the peculiar
and normalised peculiar velocities of main galaxies, and the Gaussianity
of the velocity distribution of galaxies in clusters, as estimated by the 
Anderson-Darling and Shapiro-Wilk tests. Here we see that we have clusters consisting
of up to eight components. The
peculiar velocities of the main galaxies may be large both in the clusters with one component and
with multiple components, as was seen also in  Fig.~\ref{fig:pec}. 
There are six clusters with five or more components,
but only two of them have non-Gaussian velocity distributions of galaxies
according to both tests (both are five-component clusters with 
$v_{\mathrm{pec}}  \approx 280$ km/s, so their
symbols coincide in Fig.~\ref{fig:ncompvpec}). 

\begin{figure}[ht]
\centering
\resizebox{0.45\textwidth}{!}{\includegraphics[angle=0]{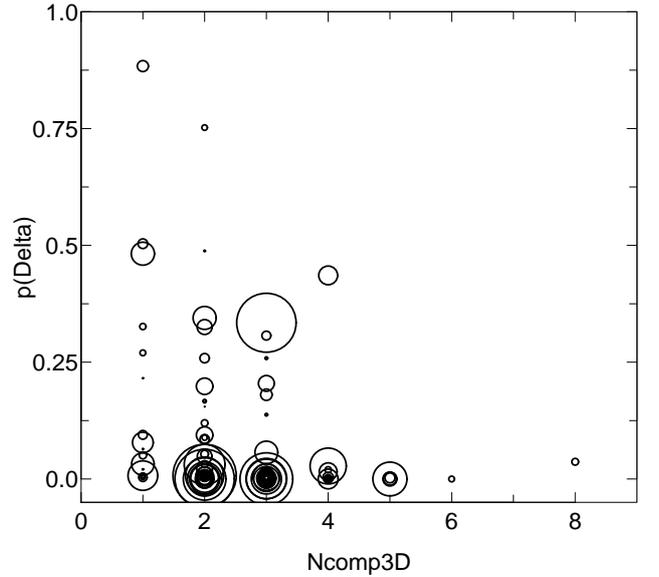}}
\caption{
3D tests: the Dressler-Shectman test $p$-value  versus the number of components 
in a cluster.
Symbol sizes are proportional to the peculiar velocities
of main galaxies, $v_{\mathrm{pec}} $. 
Small $p$-values show high probability of substructure.
}
\label{fig:delta}
\end{figure}

Figure~\ref{fig:delta} compares the results of the DS test and the 3D component 
analysis. Symbol sizes indicate the values of the peculiar velocities of the 
main galaxies in the clusters. Figure~\ref{fig:delta} shows that many 
multicomponent clusters, for which the DS test finds the presence of 
substructure, have large peculiar velocities of the main galaxes. There are 
seven one-component clusters after {\it Mclust}, for which the DS test suggests 
that these clusters have significant substructure (see also 
Table~\ref{tab:results}). Visual analysis of these clusters shows that these 
clusters have tails in their velocity distribution, and for some of these 
clusters the AD and/or SW tests also suggest a non-Gaussian distribution of 
galaxy velocities. There are 22 multicomponent clusters in our sample, which 
according to the DS test have no significant substructure. Our analysis showed 
that these clusters have components in the sky distribution of galaxies with 
similar distributions of velocities. Thus while {\it Mclust} finds better the 
components in the sky distribution of galaxies, the DS test is more sensitive to 
the possible substructure in the velocity distribution. 

\begin{figure}[ht]
\centering
\resizebox{0.45\textwidth}{!}{\includegraphics[angle=0]{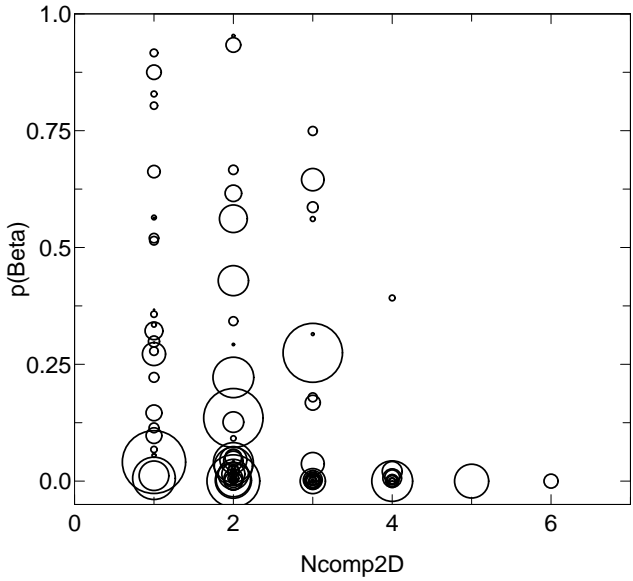}}
\caption{
2D tests: the number of cluster components versus the asymmetry ($p_{\mathrm{\beta}}$).
Symbol sizes are proportional to the peculiar velocities
of main galaxies, $V_{\mathrm{pec}}$. 
}
\label{fig:beta}
\end{figure}

Figure~\ref{fig:beta} compares the results of the $\beta$ test and the 2D 
component analysis with {\it Mclust}. The maximum number of components detected 
in the sky coordinates is  slightly smaller than that  found using both the sky 
coordinates and the velocities of galaxies. Figure~\ref{fig:beta} shows that 
there are multicomponent clusters without significant asymmetry ($p_{\mathrm{\beta}} > 0.05$) 
and with large peculiar velocities of main galaxies (30 clusters). Comparison 
with the results of the  DS test shows that 20 of these clusters have 
significant substructure, according to this test. There are 12 multicomponent 
clusters with significant asymmetry in the sky distribution of galaxies, 
according to the $\beta$ test, but without significant substructure, according 
to the DS test, showing that the velocity distribution of galaxies in different 
cluster components is similar. 

\begin{figure}[ht]
\centering
\resizebox{0.43\textwidth}{!}{\includegraphics[angle=0]{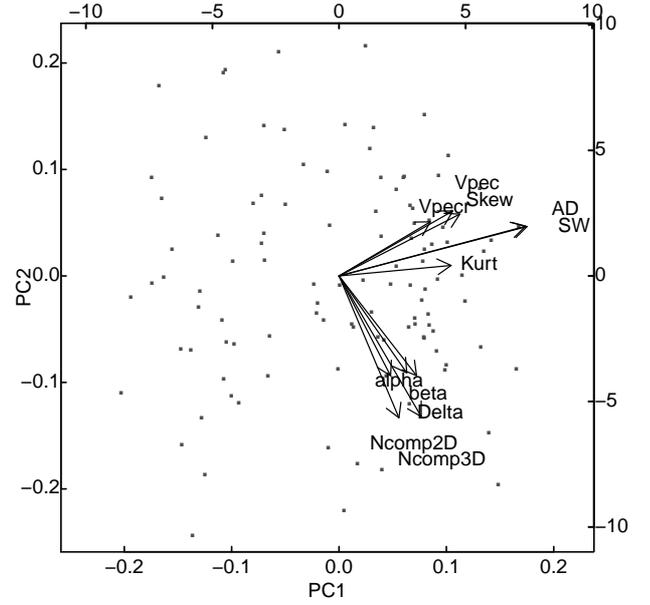}}
\caption{
Principal component analysis biplot for all tests.  
}
\label{fig:pcatests}
\end{figure}

\begin{table*}[ht]
\caption{Results of the principal component analysis for the tests.
}
\begin{tabular}{lrrrrrrrrrrr} 
\hline\hline 
                &       PC1 & PC2 & PC3 & PC4 & PC5 & PC6 & PC7 & PC8 & PC9 & PC10 & PC11  \\
\hline
$N_{\mathrm{comp3D}}$   & 0.218 &   -0.481 & -0.098 & -0.249 & -0.016 &  0.363 & -0.060 &  0.680 & -0.215 &  0.040 &  0.025 \\ 
$N_{\mathrm{comp2D}}$   & 0.158 &   -0.485 & -0.199 & -0.309 &  0.011 &  0.324 & -0.154 & -0.652 &  0.217 & -0.007 & -0.020 \\ 
$p_{\mathrm{\Delta}}$             & 0.205 &   -0.341 &  0.112 &  0.553 & -0.070 &  0.127 &  0.698 & -0.103 & -0.028 & -0.037 & -0.003 \\ 
$p_{\mathrm{\alpha}}$                & 0.136 &   -0.342 &  0.158 &  0.589 &  0.175 & -0.210 & -0.642 & -0.001 & -0.031 &  0.047 &  0.031 \\ 
$p_{\mathrm{\beta}}$                 & 0.179 &   -0.332 & -0.115 & -0.313 &  0.218 & -0.798 &  0.214 &  0.012 & -0.065 & -0.086 & -0.051 \\ 
$\log(V_{\mathrm{pec}})$& 0.301 &    0.222 & -0.544 &  0.184 &  0.101 &  0.072 & -0.071 &  0.051 & -0.001 & -0.706 & -0.076 \\ 
$V_{\mathrm{pec,r}}$    & 0.242 &    0.185 & -0.624 &  0.158 &  0.062 & -0.034 &  0.031 & -0.044 & -0.145 &  0.678 &  0.048 \\ 
$p_{\mathrm{AD}}$       & 0.499 &    0.169 &  0.195 & -0.085 &  0.018 & -0.026 &  0.013 &  0.084 &  0.438 &  0.022 &  0.689 \\ 
$p_{\mathrm{SW}}$     & 0.494 &    0.170 &  0.239 & -0.053 & -0.035 &  0.011 & -0.030 &  0.096 &  0.352 &  0.149 & -0.712 \\ 
$p_{\mathrm{kurt}}$   & 0.297 &    0.035 &  0.071 & -0.051 & -0.812 & -0.161 & -0.135 & -0.151 & -0.410 & -0.065 &  0.048 \\ 
$p_{\mathrm{skew}}$ & 0.320 &    0.211 &  0.334 & -0.136 &  0.490 &  0.177 &  0.019 & -0.234 & -0.627 & -0.022 &  0.026 \\ 
\hline
\multicolumn{3}{l}{Importance of the components} && \\ 
\hline
                &   PC1 & PC2 & PC3 & PC4 & PC5 & PC6 & PC7 & PC8 & PC9 & PC10 & PC11  \\
St. deviation     &1.783&1.392&1.283&1.090&0.969& 0.863& 0.691& 0.649& 0.524& 0.370& 0.227 \\
Prop. of Variance &0.289&0.176&0.150&0.108&0.085& 0.067& 0.043& 0.038& 0.025& 0.012& 0.004 \\
Cumulative Prop.  &0.289&0.465&0.615&0.723&0.808& 0.876& 0.919& 0.957& 0.982& 0.995& 1.000 \\
\hline

\label{tab:pcatests}  
\end{tabular}\\
\tablefoot{Last three rows present  the standard 
deviation, proportion of variance, and cumulative variance of principal 
components. 
}
\end{table*}

Next we will use the principal component analysis to compare the results of all 
tests simultaneously. We will use logarithmic scale for the peculiar velocities 
of main galaxies, this makes their range smaller. Figure~\ref{fig:pcatests} and 
Table~\ref{tab:pcatests} show the results of this analysis. In the biplot 
(Fig.~\ref{fig:pcatests}) the arrows represent the axes where each original 
variable lies, and their length is proportional to their importance within each 
PC. In calculations we use $1 - p$ instead of the $p$-value -- larger values of 
$1 - p$ suggest a higher probability to have substructure, therefore in this 
case the arrows corresponding to the number of the components in 3D and 2D point 
towards the same direction as the arrows corresponding to the DS, $\alpha$, and 
$\beta$ tests. The locations of clusters in the PC space are also plotted. In 
Table~\ref{tab:pcatests} we give the values of the principal components and  the 
standard deviations, the proportion of variance, and the cumulative variance of the
principal components. The values of the components show the importance of the 
original parameters in each PCi.

In Fig.~\ref{fig:pcatests} the arrows corresponding to the 3D and 2D tests are 
oriented towards about the same direction, the arrows corresponding to the 1D 
tests are pointed at a direction almost at the right angle to this, and form 
another set. This suggests that the results of the 3D and 2D tests are 
correlated, as well as the results of 1D tests among themselves. Among the 3D 
and 2D tests the highest importance in first two principal PCA components have 
the numbers of components determined with normal mixture modelling, followed 
by the DS test. The $\alpha$  and $\beta$ tests   are of almost equal 
importance. Among the 1D tests, the results of the Anderson-Darling and the 
Shapiro-Wilk tests have equal importance in determining the non-Gaussianity of 
the velocity distribution.

Table~\ref{tab:pcatests} shows that seven principal components are needed to 
explain more than 90\% of  the variance in the parameters, thus there is no one 
dominant test which could find most of the information about multimodality of the
galaxy distribution in clusters.

Different types of clusters populate different regions in the PC1-PC2 plane. 
Unimodal clusters with Gaussian distribution of velocities are located in the 
upper lefthand part of the plot and have larger PC2 and larger negative PC1 
values (for example, the clusters 608 and 58604). Multimodal clusters populate 
the lower righthand area of the biplot (the clusters 34727 and 914). Clusters 
with small values of the peculiar velocities of main galaxies, with a Gaussian 
distribution of velocities, but with a large number of components populate the 
lefthand lower area of the PC1-PC2 plane (the clusters 29348 and 4744). 
Multimodal clusters with large peculiar velocities of main galaxies and with 
non-Gaussian velocity distributions are located in the upper righthand area of 
the plane (the clusters 4122, 23374 and others).

\begin{figure*}[ht]
\centering
\resizebox{0.90\textwidth}{!}{\includegraphics[angle=0]{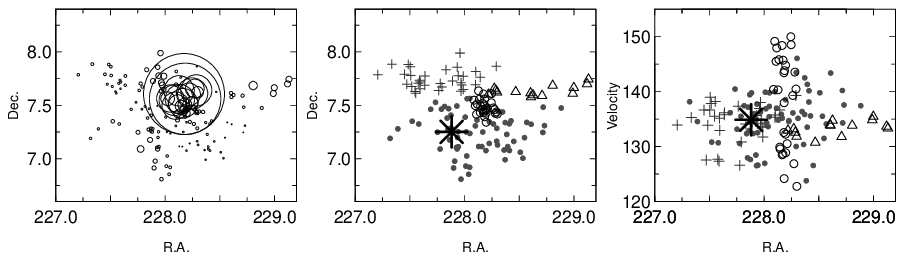}}\\
\resizebox{0.90\textwidth}{!}{\includegraphics[angle=0]{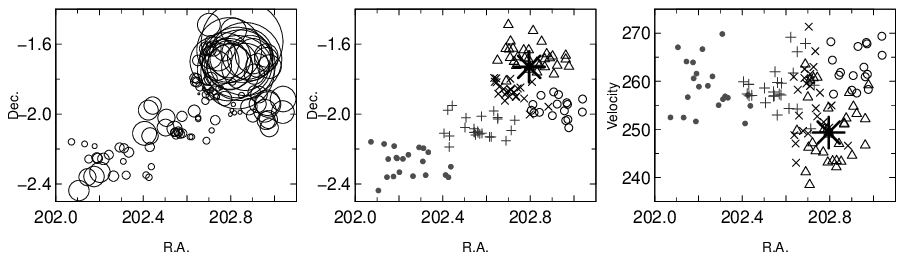}}
\caption{
Two of the most non-Gaussian clusters in our sample.
Upper row: the cluster 34726 (Abell cluster A2028). Lower row: the cluster 914 (A1750).
From left to right: the DS test bubble plot (here symbol sizes are proportional to the
$e^\delta$), and R.A. vs. Dec., and R.A. vs. velocity (in $10^{2}km/sec$) plots; 
the symbols show different components as found by {\it Mclust}. 
The star marks the location of the main galaxy.
}
\label{fig:grnonG}
\end{figure*}

\begin{figure*}[ht]
\centering
\resizebox{0.90\textwidth}{!}{\includegraphics[angle=0]{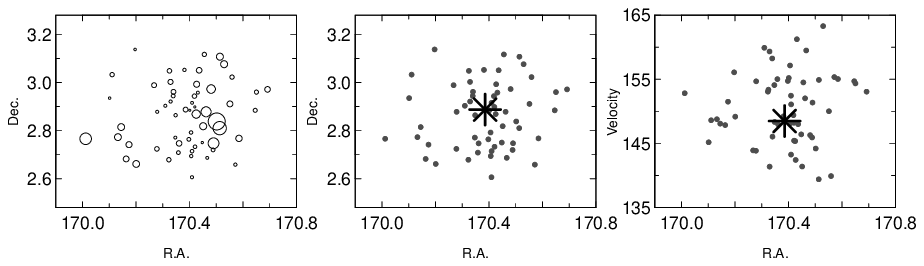}}\\
\resizebox{0.90\textwidth}{!}{\includegraphics[angle=0]{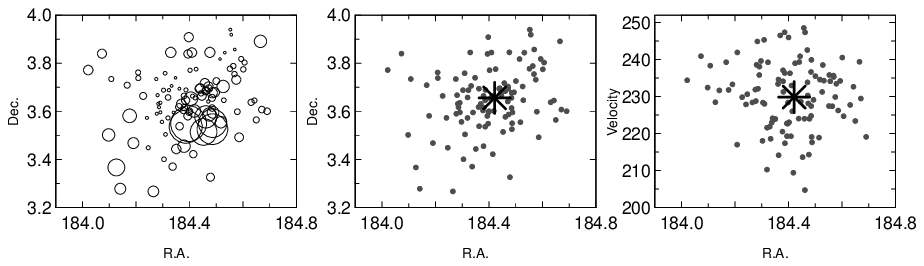}}
\caption{
Two of the most Gaussian clusters in our sample.
Upper row: the cluster 58604. Lower row: the cluster 60539.
From left to right: the DS test bubble plot
(symbol sizes are proportional to the
$e^\delta$), and R.A. vs. Dec., and R.A. vs velocity (in $10^{2}km/s$) plots; 
the symbols show different components as found by {\it Mclust}. 
The star marks the location of the main galaxy.
}
\label{fig:grG}
\end{figure*}

\subsection{Examples of multimodal and unimodal clusters}

In our sample there are eight clusters which show signatures of  
non-Gaussianity in their galaxy distribution by all the tests. They have multiple 
components, the 3D and 2D tests confirm the presence of substructures and of asymmetry, 
and their velocity distributions differ from Gaussian at a very high statistical 
significance: the clusters 880,  4122, 28387, 34726, 58305, 67297, 68625, and 73088. Many 
clusters with multiple components are missing from this list, for example the
cluster 914, a well-known binary merging Abell cluster A1750 in the Sloan Great 
Wall \citep[][and references therein]{2010A&A...522A..92E, 2011ApJ...736...51E, 
2011A&A...532A...5E}, because in many multicomponent clusters galaxies from 
different components have similar distributions of velocities.

Figure~\ref{fig:grnonG} shows the distribution of galaxies in a cluster with four
components, the cluster 34726, and in the cluster 914 with five components.

The cluster 34726 is one of the two clusters in our sample which can be 
identified with the Abell cluster A2028 in sky projection, another one is the 
cluster 34727. A2028 is a merging X-ray cluster with substructure 
\citep{2006A&A...450....9F, 2010A&A...522A..34G}, which redshift close to the 
redshift of our cluster 34727. This is one example of projections in the Abell 
cluster catalogue. Figure~\ref{fig:grnonG} (upper row) shows that the sky 
distribution of galaxies in the cluster 34726 has two main components. In one of 
them the concentration of galaxies is especially high. This component is well 
seen in the distribution of velocities where very large velocity dispersion 
suggest a finger-of-god effect (right panel). Here also according to the DS test 
the probability of substructure is high. The $\alpha$  and $\beta$ tests confirm 
the presence of substructure and asymmetry in the distribution of galaxies, the 
AD and SW tests show non-Gaussianity of the distribution of velocities of 
galaxies in this cluster. The peculiar velocity of the main galaxy is small.

The cluster 914 (Fig.~\ref{fig:grnonG}, lower row) that corresponds to the Abell 
cluster A1750 is a well-known merging binary X-ray cluster \citep{don01, 
belsole04, bur04, hwanglee09, 2010A&A...522A..92E}. In this cluster we see even 
five components. The DS test shows that the probability of substructure is the 
largest in the component where the main galaxy is located 
(Fig.~\ref{fig:grnonG}, lower left panel). The peculiar velocity of the main 
galaxy is large. Other 3D and 2D tests confirm the presence of substructure and 
an asymmetrical galaxy distribution in A1750. Galaxies from different components 
in this cluster have similar velocity distributions, so according to the AD and 
SW tests their distribution is Gaussian \citep[for details and references we 
refer also to ][]{2010A&A...522A..92E}.

Next we list  the clusters for which all tests confirm unimodality and the 
Gaussian velocity distribution of galaxies: 608,  5217, 25078, 39914, 50129, 
58604, and 67116. In Fig.~\ref{fig:grG} we plot the distribution of galaxies in 
two of them -- in the cluster 58604, and in the cluster  60539 \citep[A1516, the 
richest cluster in the supercluster SCl~111 in the Sloan Great Wall, described 
in][]{2010A&A...522A..92E}. For the cluster 58604 all our tests confirm that 
this is an unimodal cluster. The peculiar velocity of the main galaxy is rather 
large for an unimodal cluster, about 200 km/s. The cluster A1516 has the most 
regular ellipsoidal sky distribution of galaxies among the clusters in our 
sample, with a small peculiar velocity of the main galaxy. However, the 
distribution of velocities in the central part of the cluster shows an excess of 
negative (in respect of the cluster center) velocities, and according to the DS 
and $\alpha$ tests this cluster may have substructure \citep[see also an 
analysis of this cluster in][]{2010A&A...522A..92E}. This may be a signature of 
a line-of-sight merger \citep[see also][]{1996ApJS..104....1P}.  

\begin{table*}[ht]
\caption{Results of the principal component analysis for the test results and
for the physical parameters of clusters, combined.
}
\begin{tabular}{lrrrrrrr} 
\hline\hline 
                &       PC1 & PC2 & PC3 & PC4 & PC5 & PC6 & PC7  \\
\hline
$N_{\mathrm{comp3D}}$        & 0.218 & -0.596 &  0.407 & -0.115 &  0.600 &  0.232 & -0.050 \\ 
$\log(V_{\mathrm{pec}})$     & 0.185 &  0.305 &  0.415 & -0.811 & -0.147 & -0.133 &  0.039 \\ 
$p_{\mathrm{\Delta}}$                  & 0.284 & -0.171 &  0.602 &  0.427 & -0.571 & -0.051 &  0.118 \\ 
$\log(L_{\mathrm{tot}})$     & 0.537 &  0.116 & -0.287 & -0.010 &  0.035 &  0.351 &  0.699 \\ 
$\log(N_{\mathrm{gal}})$     & 0.509 & -0.004 & -0.153 &  0.120 &  0.242 & -0.799 & -0.063 \\ 
$\log(rms~velocity)$         & 0.355 &  0.603 &  0.162 &  0.268 &  0.236 &  0.343 & -0.487 \\ 
$\log(r_{\mathrm{vir}})$     & 0.401 & -0.379 & -0.406 & -0.242 & -0.418 &  0.211 & -0.500 \\ 
\hline
\multicolumn{3}{l}{Importance of components} && \\ 
\hline
                  &  PC1   & PC2    & PC3    & PC4    & PC5   & PC6 & PC7    \\
St. deviation     & 1.627  & 1.115  & 1.014  & 0.950  & 0.795 &  0.609  & 0.418   \\
Prop. of Variance & 0.378  & 0.177  & 0.147  & 0.129  & 0.090 &  0.053  & 0.025   \\
Cumulative Prop.  & 0.378  & 0.556  & 0.703  & 0.832  & 0.922 &  0.975  & 1.000   \\
\hline

\label{tab:pcaphys}  
\end{tabular}\\
\end{table*}

In Tables~\ref{tab:unicl} -- ~\ref{tab:multiresults} 
we present the data about the most unimodal and most multimodal 
clusters, as well as data about the clusters with at least 
five components, and about those one-component clusters for which the DS test
found significant substructure.

\subsection{Non-Gaussianity parameters and physical properties of clusters}

\begin{figure}[ht]
\centering
\resizebox{0.43\textwidth}{!}{\includegraphics[angle=0]{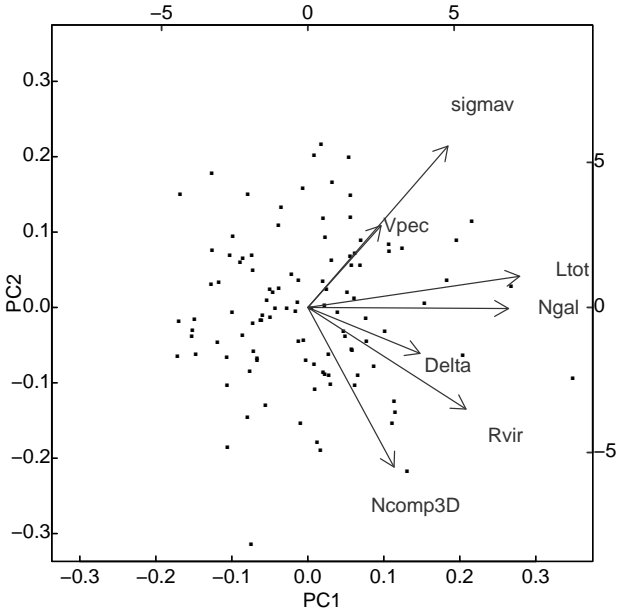}}
\caption{
Principal component analysis with $N_{\mathrm{comp3D}}$, 
$V_{\mathrm{pec}}$, and $p_{\mathrm{\Delta}}$, and the physical parameters of clusters
($L_{tot}$, $N_{gal}$, $r_{vir}$, and $rms~velocity$). 
}
\label{fig:pcaphys}
\end{figure}

The principal component analysis can help us to study the relation between the 
physical parameters of clusters and the presence of substructure. In these 
calculations we include the number of components as determined with the 3D 
normal mixture modelling, the $p$-value for the DS test, $p_{\mathrm{\Delta}}$, 
the peculiar velocity of the main galaxy in a cluster, $V_{\mathrm{pec}}$, the 
number of galaxies in a cluster, $N_{\mathrm{gal}}$, the total luminosity of a 
cluster, $L_{\mathrm{tot}}$, the virial radius of a cluster, $r_{\mathrm{vir}}$, 
and the rms velocity of the cluster galaxies. We again use 
$1 - p_{\mathrm{\Delta}}$, and use logarithms of the peculiar velocities of main 
galaxies and of the physical parameters. Figure~\ref{fig:pcaphys} and 
Table~\ref{tab:pcaphys} show the results of this analysis.

In Fig.~\ref{fig:pcaphys} the arrows corresponding to the tests for 
substructure and the arrows corresponding to the physical parameters of clusters 
do not point to the same direction -- the correlations between the parameters 
are not strong, when we consider all parameters simultaneously. The coefficients of the 
first principal component  in Table~\ref{tab:pcaphys} show that the physical 
parameters are of larger importance than the substructure indicators. The 
coefficient corresponding to the luminosity of a cluster is the largest, but the 
differences between the values of the coefficients are not large, so there is no 
single dominant parameter. Among the coefficients of the second principal 
component the number of components has the largest (absolute) coefficient. Among 
the coefficients of the third principal component the peculiar velocities of 
the main galaxy, and the rms velocity of galaxies in a cluster are of equal importance. 
In Fig.~\ref{fig:pcaphys} the arrow corresponding to the peculiar velocity of 
the main galaxy in a cluster points to the same direction as the arrow 
for the rms velocity of a cluster, showing that these velocities are 
larger in clusters with larger rms velocities. These clusters are also richer.  

Table~\ref{tab:pcaphys} shows that the first principal component accounts for 
almost 38\% of the variance of parameters. Five principal components are needed 
to explain more than 90\% of the variance of the parameters -- clusters are 
complicated objects, the properties of which cannot be explained with a small 
number of parameters \citep[in contrast to superclusters which can be described 
with a small number of parameters, see][]{2011A&A...535A..36E}. 

In the PC1-PC2 plane
one-component clusters with a Gaussian distribution of velocities are located at 
the upper lefthand part of the plot and have larger PC2 and smaller  (larger 
negative) PC1 values (for example, the clusters 608 and 28508). Multicomponent 
clusters of high luminosity populate the lower and middle righthand area of the 
biplot (the clusters 34727 and 914). Less luminous multicomponent clusters 
populate the lefthand lower area of the PC1- PC2 plane (the clusters 11474 and 
11015). One-component clusters with large rms velocities of galaxies populate 
the upper righthand area of the plane (the clusters 16350, 17210, and others).

We checked for pairwise correlations between the physical parameters of clusters 
and the substructure characteristics and found that correlations between the number 
of galaxies in clusters, the total luminosity of clusters, the virial radius 
of clusters, and the numbers of components and  the presence of substructure
according to the 3D and 2D tests are statistically highly significant --
richer, larger and more luminous clusters have a larger amount of substructure.

\section{Discussion and conclusions}
\label{sect:discussion} 

\subsection{Comparison with other studies}
\label{sect:comp} 

An increasing number of studies have shown recently the presence of substructure 
in groups and clusters of galaxies determined using optical or X-ray data (see
references in Sect.~\ref{sect:intro}). We detected multiple components in 
more than 80\% of the clusters using multidimensional normal mixture modelling. 
Comparison with other studies in Sect.~\ref{sect:results} showed that although 
this is a higher fraction of multimodal clusters than found in other studies, 
comparison of the fractions of clusters with substructure or with non-Gaussian 
velocity distributions determined with similar methods gives the results in a 
good agreement with others. 
This supports our choice of the parameters for the FoF algorithm for group 
definition and suggests that  in our catalogue groups  with substructure are 
real groups and not complexes of small groups, artificially linked together by an 
unreasonable choise of linking parameters. 

The fractions of clusters with substructure found in X-ray clusters by 
\citet{2007A&A...470...39R}  is almost the same that we found with the 2D normal 
mixture modelling. In these studies there are eight common clusters; in seven of 
them substructure have been detected in both studies. 
\citet{2007A&A...470...39R} found two components in A602 (cluster 43545), this 
is unimodal cluster according to our calculations. In the sky distribution this 
cluster has a very elongated shape; this may be visible in X-rays as two 
components. 

Comparison with \citet{2006A&A...450....9F} shows that they detected 
substructures  in all 14 clusters common for both studies; we found multiple 
components in 12 clusters.  A1809 (our cluster 67116) is an unimodal cluster according 
to {\it Mclust}, although the DS and $\alpha$ tests detected substructure at the 90\% 
significance level. We detected no substructure in A1650 (25078) while 
\citet{2006A&A...450....9F} mention that there is one group of galaxies in the 
field of the galaxies in this cluster. Therefore, an overall agreement 
between the results of our study and \citet{2006A&A...450....9F} is very good, 
considering that the sample selection and the methods to analyse modality of 
clusters are different.

The cluster A1750 (our cluster 914) has been studied for substructure by other authors, too. We 
mentioned above that the components found in this cluster by us coincide well with 
those found in other studies \citep[see][for details]{2010A&A...522A..92E}. 
We found an especially good agreement between our 
results and those by \citet{bur04} and \citet{hwanglee09}. 

We have three common clusters with those searched for rotation by 
\citet{2007ApJ...662..236H}. We found that all of them (A1035, A1139, and A2169; 
these are our clusters 20159, 3714, and 12508) are multimodal clusters which 
agrees with the results of the DS test by \citet{2007ApJ...662..236H}. They also 
suggest that the clusters A1139 and A2169 may be rotating or merging. We have 
also five common clusters with those studied by \citet{bur04} using data from 
the 2dFGRS. All of them (A1139, A1238, A1620, A1663, and A1750, these are our 
clusters 3714, 4744, 24554, 24829, and 914) are multimodal clusters, as found in 
both studies. Eight of ten common clusters with those studied by 
\citet{oegerlehill2001}  are multimodal in our study, for one unimodal cluster 
(A2089, our cluster 39914) the $p_{\mathrm{\Delta}} = 0.051$ that shows the 
presence of substructure, as found in \citet{oegerlehill2001}, but not at a very 
high significance level.

We showed that  the peculiar velocities of main galaxies in a large fraction of 
clusters are  large, and the most luminous galaxy (cluster main galaxy) is often 
not located close to the cluster centre. 
{\it
This means that the most luminous 
galaxies in clusters are not the central galaxies. 
}
Actually, there is no 
stringent reason for that. In the merger scenario, the clusters form through 
merging of smaller groups/clusters. The main galaxy in a new cluster is one of 
the main galaxies of the merged clusters. As pointed out in \citet{tempel09}, 
the second luminous galaxies in rich clusters have been main galaxies before the 
latest merger event between the clusters. In this study, we analyse the richest 
clusters in the SDSS sample, and such rich clusters form through merging of 
other clusters. The presence of substructure in these clusters supports this 
idea.  For the main galaxies this means that the main galaxies in rich clusters have 
not yet found their place in the cluster centre, they are still located close to 
the centres of their parent clusters (subclusters of the new cluster). Large 
distances from the cluster centres and large peculiar velocities of the 
brightest galaxies in clusters indicate that clusters are not yet fully virialized 
after the last merger. Similar results were obtained by 
\citet{2010A&A...522A..92E, 2011ApJ...736...51E} in the study of groups in the 
Sloan Great Wall.

Cosmological simulations show the merging and growth of dark matter haloes 
\citep[][and references therein]{1992ApJ...393..477R, 2002MNRAS.336..112M, 
2008MNRAS.388.1537M, 2010MNRAS.406.2267F, 2011MNRAS.tmp.1734P}. The late time 
formation of the main haloes and the number of recent major mergers can cause 
the late time subgrouping of haloes \citep{2008ApJ...682L..73S, 
2010A&A...522A..92E, 2011MNRAS.tmp.1734P}.

The principal component analysis of the substructure indicators suggests that the 
results of the 3D and 2D tests are correlated, as well as the results of the 1D tests. 
Among the 3D and 2D tests the highest importance have the 3D normal mixture 
modelling and the DS test. With the 3D tests we detected the largest fraction of 
clusters with substructure, followed by the 2D tests. The PCA showed that there is no 
single dominant test which could find most of the information about multimodality 
of the galaxy distribution in clusters. Also \citet{1996ApJS..104....1P} 
concluded that the higher the dimensionality of the test, the higher is the 
probability to detect substructure. However, the tests are sensitive to the 
different aspects of substructure and to non-Gaussianity, thus it is preferable 
to use several tests of different dimensionality.

{\it
Richer, larger and more luminous clusters have a larger amount of 
substructure. 
}
The principal component analysis using both the substructure indicators and 
the physical parameters of clusters showed that five principal components are needed 
to explain more than 90\% of variance in data --  galaxy clusters are 
complicated objects, the properties of which cannot be explained with a small number of 
parameters, as was shown recently also by \citet{2011MNRAS.415L..69J} and 
\citet{2011MNRAS.416.2388S} by the PCA analysis of dark matter haloes.

Cosmological simulations show that the fraction of substructure is higher in 
high-redshift haloes \citep{2010MNRAS.404..502G}. \citet{2007MNRAS.379.1343S} 
and    ~\citet{2008ApJ...684..933G} found evidence of  substructure in clusters 
of high-redshift superclusters at $z \approx 0.9$. More data about high-redshift 
clusters is needed to make a statistical comparison.

Current theories of the formation and evolution of galaxies, and galaxy groups 
and clusters tell that galaxies and their systems form in virialised dark matter 
haloes \citep{1978MNRAS.183..341W, 1991ApJ...379...52W}, which grow 
hierarchically by merging of smaller mass haloes \citep[see, e.g.,][]{loeb2008}. 
The halo model of group and cluster properties and clustering has been 
successful in explaining several properties of groups and clusters, their galaxy 
population, and clustering \citep[][and references therein]{2002ApJ...575..587B, 
2006astro.ph.10524B, 2007MNRAS.376..841V, 2009MNRAS.392.1467S}. The halo model 
has been also used in studies of the structure of superclusters 
\citep{2011ApJ...736...51E}. In virialised clusters  galaxies follow the 
gravitational potential and the main galaxies of clusters (their brightest 
galaxies) lie at the centers of clusters and have small peculiar velocities 
\citep{ost75,merritt84,malumuth92, 2002ApJ...575..587B, 2005MNRAS.356.1293Y}. 
The most important halo parameter is the halo mass, although several studies 
have shown that other parameters are also important in shaping the properties of 
dark matter haloes, like the formation time, concentration and others 
\citep[][and references therein]{2011MNRAS.415L..69J, 2011MNRAS.416.2388S, 
2011arXiv1109.4169C}.

We showed that  a significant fraction of the main   galaxies in clusters have 
large peculiar velocities, and 
{\it
more than 80\% of clusters in our sample have 
substructure and/or non-Gaussian velocity distributions. 
}
\citet{2000A&A...354..761K} applied the DS statistics to search for substructure 
of clusters in numerical simulations. They studied clusters from a set of 
cosmological models and found that in all models, the DS test detected 
significant substructure in 30-40 \% of clusters, which is about half of what we 
found in this study. The high fraction of clusters with substructure and large 
peculiar velocities of main galaxies  show that clusters are not in dynamical 
equilibrium. We also found that galaxy clusters are complicated objects, the 
properties of which cannot be explained with a small number of parameters. These 
differences between observations and simulations do not fit well into the halo 
model framework and have yet to be explained.

\subsection{Conclusions}
\label{sect:conclusions} 

We searched for substructure and non-Gaussian velocity distributions
in rich clusters drawn from the SDSS DR8. 
We present lists of unimodal and multimodal clusters.
Our conclusions are as follows.

\begin{itemize}
\item[1)]
We showed, using a number of tests, that more than 80\% of 
rich clusters  have substructure
and/or non-Gaussian velocity distributions of galaxies. 
\item[2)] 
The peculiar velocities of the main galaxies in clusters, and their distances 
from the cluster centre are large, especially in clusters with multiple 
components. In multicomponent clusters the brightest galaxies are typically 
located near the centre of one of the components.
\item[3)] 
The  largest number of clusters with substructure
was detected by multidimensional normal mixture modelling, followed
by the Dressler-Shectman test.  
\item[4)]
Principal component analysis shows that there is no one 
dominant test which could find most of the information about the multimodality of 
galaxy distribution in clusters.
Different tests are sensitive to different signatures of multimodality,
therefore it is important to use several tests to search for substructure
in clusters.  
\item[5)]
Richer, larger, and more luminous clusters with larger rms velocities
have larger amount of substructure and 
larger (normalised) peculiar velocities of the main galaxies. 
\item[6)]
Principal component analysis using both substructure indicators and the physical
parameters of clusters showed that galaxy clusters are complicated 
objects, the properties of which cannot be explained with a small number of parameters. 
\item[7)]
Our results show that simple halo model do not explain all the properties of 
observed clusters. 
The halo model assumes that haloes (clusters) are virialised, while
we found that they are not. Also, the fraction of observed clusters
with substructure is larger than that found in simulations.
\end{itemize}

The presence of substructure, large distances of main galaxies from the cluster 
centre, and their large peculiar velocities is a sign of mergers and/or infall, 
and suggests that most clusters in our sample are not yet in dynamical 
equilibrium. The high frequency of such clusters tells  that mergers between 
groups and clusters are common -- galaxy groups continue to grow and are still 
assembling. Our unimodal clusters are examples of clusters which are
probably already in dynamical equilibrium and therefore can be used in the studies
of cluster characteristics like mass estimation, analysis of concentration
and others reliably.

To understand better the properties of galaxy clusters and hence the formation 
and evolution  of the structures in the Universe we plan to analyse in detail 
galaxy populations in different cluster components, and the connection of the 
properties of clusters and the environment where they reside.

\section*{Acknowledgments}

We thank our referee for  useful suggestions which helped to improve the paper. 

We are pleased to thank the SDSS Team for the publicly available data
releases.  
Funding for the Sloan Digital Sky Survey (SDSS) and SDSS-II has been
  provided by the Alfred P. Sloan Foundation, the Participating Institutions,
  the National Science Foundation, the U.S.  Department of Energy, the
  National Aeronautics and Space Administration, the Japanese Monbukagakusho,
  and the Max Planck Society, and the Higher Education Funding Council for
  England.  The SDSS Web site is \texttt{http://www.sdss.org/}.
  The SDSS is managed by the Astrophysical Research Consortium (ARC) for the
  Participating Institutions.  The Participating Institutions are the American
  Museum of Natural History, Astrophysical Institute Potsdam, University of
  Basel, University of Cambridge, Case Western Reserve University, The
  University of Chicago, Drexel University, Fermilab, the Institute for
  Advanced Study, the Japan Participation Group, The Johns Hopkins University,
  the Joint Institute for Nuclear Astrophysics, the Kavli Institute for
  Particle Astrophysics and Cosmology, the Korean Scientist Group, the Chinese
  Academy of Sciences (LAMOST), Los Alamos National Laboratory, the
  Max-Planck-Institute for Astronomy (MPIA), the Max-Planck-Institute for
  Astrophysics (MPA), New Mexico State University, Ohio State University,
  University of Pittsburgh, University of Portsmouth, Princeton University,
  the United States Naval Observatory, and the University of Washington.

The present study was supported by the Estonian Science Foundation
grants No. 8005, 7765, and MJD272, by the Estonian Ministry for Education and
Science research project SF0060067s08, and by the European Structural Funds
grant for the Centre of Excellence "Dark Matter in (Astro)particle Physics and
Cosmology" TK120. This work has also been supported by
ICRAnet through a professorship for Jaan Einasto.
P.N. was supported by the Academy of Finland, P.H. by Turku University Foundation. 
V. M. was supported by the Spanish MICINN CONSOLIDER projects 
ATA2006-14056 and CSD2007-00060, including FEDER contributions, 
and by the Generalitat Valenciana project of excellence PROMETEO/2009/064.

\bibliographystyle{aa}
\bibliography{sub.bib}

\begin{appendix}
\section{Data on clusters} 
\label{sec:cltable}

\begin{table*}[ht]
\caption{Data on unimodal clusters}
\begin{tabular}{rrrrrrrrr} 
\hline\hline  
(1)&(2)&(3)&(4)&(5)& (6)&(7)&(8)&(9) \\      
\hline 
 ID&$N_{\mathrm{gal}}$& $\mathrm{R.A.}$ & $\mathrm{Dec.}$ &$\mathrm{Dist.}$ & $L_{\mathrm{tot}}$ & $\sigma$ & $r_{\mathrm{vir}}$ & Abell ID \\
 & &[deg]&[deg]&[$h^{-1}$ Mpc]&[$10^{10} h^{-2} L_{\sun}$]&$[km s^{-1}$]&[$h^{-1}$ Mpc]& \\
\hline
  608 &  60 & 245.23 & 29.83 & 284.84 & 132.16 &  532.83 & 0.66 & \object{A 2175}\\
  748 &  79 & 159.85 &  5.22 & 206.78 &  93.59 &  748.24 & 0.43 & \object{A 1066}\\
 5217 &  89 & 180.06 & 56.22 & 193.33 &  94.88 &  577.08 & 0.61 & \object{A 1436}\\
25078 &  51 & 194.69 & -1.70 & 249.24 &  87.64 &  498.42 & 0.61 & \object{A 1650}\\
28272 &  51 & 248.33 & 11.80 & 153.80 &  40.89 &  355.14 & 0.50 & ---\\
32663 &  51 & 225.59 & 21.36 & 186.20 &  61.37 &  435.82 & 0.73 & ---\\
39914 &  63 & 233.12 & 28.02 & 218.38 &  69.77 &  446.50 & 0.65 & \object{A 2089}\\
43336 &  68 & 207.98 & 46.33 & 186.81 &  73.08 &  471.40 & 0.60 & ---\\
43545 &  51 & 118.25 & 29.41 & 179.41 &  42.75 &  577.85 & 0.51 & \object{A 602}\\
50129 &  52 & 205.43 & 26.39 & 225.02 &  61.83 &  449.92 & 0.51 & \object{A 1775}\\
58604 &  58 & 170.39 &  2.86 & 150.50 &  39.84 &  528.86 & 0.42 & --- \\
60539 & 107 & 184.40 &  3.65 & 230.10 & 136.74 &  830.47 & 0.55 & \object{A 1516}\\
63361 &  72 & 205.59 &  2.26 & 230.57 & 103.06 &  646.10 & 0.64 & \object{A 1773}\\
67116 &  80 & 208.29 &  5.17 & 236.99 & 114.14 &  651.61 & 0.44 & \object{A 1809}\\

\label{tab:unicl}  
\end{tabular}\\
\tablefoot{                                                                                 
Columns are as follows:
1: ID of the cluster;
2: the number of galaxies in the cluster, $N_{\mathrm{gal}}$;
3--4: cluster right ascension and declination;
5: cluster comoving distance;
6: cluster total luminosity;
7: rms velocity of galaxies in the cluster;
8: cluster virial radius;
9: Abell ID of the cluster.
}
\end{table*}

\begin{table*}[ht]
\caption{Results of the tests. Unimodal clusters}
\begin{tabular}{rrrrrrccccccccc} 
\hline\hline  
(1)&(2)&(3)&(4)&(5)& (6)&(7)&(8)&(9)&(10)&(11)&(12)& (13)&(14)&(15) \\      
\hline 
 ID   &$N_{\mathrm{gal}}$ & $v_{\mathrm{pec}}$ & $v_{\mathrm{pec,r}}$ &
$N_{\mathrm{c3D}}$ & $un3D$ & $p_{\mathrm{\Delta}}$ & $p_{\mathrm{\alpha}}$ & $N_{\mathrm{c2D}}$ & $un2D$ &
$p_{\mathrm{\beta}}$ & $p_{\mathrm{AD}}$ & $p_{\mathrm{SW}}$ & $p_{\mathrm{kurt}}$ &  $p_{\mathrm{skew}}$  \\
\hline
  608 &  60 &  -228.43 & -0.429 & 1 & $1\cdot10^{-3}$ & 0.884 & 0.080 & 1 & $1\cdot10^{-3}$ & 0.299 & 0.735 & 0.860 & 0.611 & 0.511 \\
  748 &  79 &   613.95 &  0.821 & 1 & $1\cdot10^{-3}$ & 0.007 & 0.007 & 1 & $1\cdot10^{-3}$ & 0.011 & 0.308 & 0.076 & 0.154 & 0.564 \\
 5217 &  89 &   -19.11 & -0.033 & 1 & $1\cdot10^{-3}$ & 0.216 & 0.110 & 1 & $1\cdot10^{-3}$ & 0.367 & 0.240 & 0.309 & 0.136 & 0.658 \\
25078 &  51 &  -120.45 & -0.242 & 1 & $1\cdot10^{-3}$ & 0.270 & 0.378 & 1 & $1\cdot10^{-3}$ & 0.829 & 0.258 & 0.513 & 0.753 & 0.452 \\
28272 &  51 &   464.63 &  1.308 & 1 & $1\cdot10^{-3}$ & 0.032 & 0.213 & 3 & $1\cdot10^{-3}$ & 0.645 & 0.244 & 0.307 & 0.568 & 0.298 \\
32663 &  51 &    96.03 &  0.220 & 1 & $1\cdot10^{-3}$ & 0.008 & 0.147 & 1 & $1\cdot10^{-3}$ & 0.564 & 0.060 & 0.041 & 0.174 & 0.118 \\
39914 &  63 &   148.99 &  0.334 & 1 & $1\cdot10^{-3}$ & 0.051 & 0.106 & 1 & $1\cdot10^{-3}$ & 0.804 & 0.904 & 0.863 & 0.566 & 0.759 \\
43336 &  68 &   -97.43 & -0.207 & 1 & $1\cdot10^{-3}$ & 0.003 & $1\cdot10^{-3}$ & 1 & $1\cdot10^{-3}$ & 0.335 & 0.980 & 0.941 & 0.756 & 0.782 \\
43545 &  51 &    81.55 &  0.141 & 1 & $1\cdot10^{-3}$ & 0.004 & 0.011 & 1 & $1\cdot10^{-3}$ & 0.055 & 0.047 & 0.200 & 0.817 & 0.369 \\
50129 &  52 &  -129.26 & -0.287 & 1 & $1\cdot10^{-3}$ & 0.326 & 0.737 & 1 & $1\cdot10^{-3}$ & 0.357 & 0.446 & 0.421 & 0.181 & 0.770 \\
58604 &  58 &  -201.00 & -0.380 & 1 & $1\cdot10^{-3}$ & 0.504 & 0.232 & 1 & $1\cdot10^{-3}$ & 0.222 & 0.522 & 0.677 & 0.363 & 0.847 \\
60539 & 107 &   -24.93 & -0.030 & 1 & $1\cdot10^{-3}$ & 0.021 & 0.002 & 1 & $1\cdot10^{-3}$ & 0.563 & 0.776 & 0.645 & 0.882 & 0.479 \\
63361 &  72 &  -175.06 & -0.271 & 1 & $1\cdot10^{-3}$ & 0.004 & 0.179 & 1 & $1\cdot10^{-3}$ & 0.278 & 0.240 & 0.455 & 0.347 & 0.820 \\
67116 &  80 &  -183.37 & -0.281 & 1 & $1\cdot10^{-3}$ & 0.094 & 0.089 & 2 & 0.149 & 0.342 & 0.929 & 0.805 & 0.930 & 0.862 \\

\label{tab:uniresults}  
\end{tabular}\\
\tablefoot{                                                                                 
Columns are as follows:
1: ID of a cluster;
2: the number of galaxies in a cluster, $N_{\mathrm{gal}}$;
3: the peculiar velocity of the main galaxy ($km s^{-1}$),
4: the normalised peculiar velocity of the main galaxy,
$v_{\mathrm{pec}}$/$\sigma_{\mathrm{v}}$;
5: the number of components in 3D in a cluster, $N_{\mathrm{comp}}$;
6: the median uncertainty of classification  (hereafter $1\cdot10^{-3}$ denotes
the value $<1\cdot10^{-3}$);
7: the $p$-value for the DS test;
8: the $p$-value for the $\alpha$ test;
9: the number of components in 2D in a cluster, $N_{\mathrm{comp}}$;
10: the median uncertainty of classification in 2D;
11: the $p$-value for the $\beta$ test;
12: the $p$-value for the AD test;
13: the $p$-value of the SW test;
14: the $p$-value for the kurtosis test;
14: the $p$-value for the skewness test.
}
\end{table*}

\begin{table*}[ht]
\caption{Data on multimodal clusters}
\begin{tabular}{rrrrrrrrr} 
\hline\hline  
(1)&(2)&(3)&(4)&(5)& (6)&(7)&(8)&(9) \\      
\hline 
 ID&$N_{\mathrm{gal}}$& $\mathrm{R.A.}$ & $\mathrm{Dec.}$ &$\mathrm{Dist.}$ & $L_{\mathrm{tot}}$ & $\sigma$ & $r_{\mathrm{vir}}$ & Abell ID \\
 & &[deg]&[deg]&[$h^{-1}$ Mpc]&[$10^{10} h^{-2} L_{\sun}$]&$[km s^{-1}$]&[$h^{-1}$ Mpc]& \\
\hline
  880 &  57 & 211.40 &  6.27 & 249.48 & 101.82 &  411.37 & 0.84   & --- \\
  914 & 119 & 202.65 & -1.94 & 256.38 & 227.45 &  657.41 & 0.83   & \object{A 1750} \\
 4122 &  88 & 173.06 & 56.09 & 159.40 &  68.62 &  963.42 & 0.49   & \object{A 1291} \\
11015 &  52 & 212.51 & 55.07 & 124.69 &  45.80 &  303.10 & 0.46   & --- \\
11474 &  51 & 218.27 & 52.90 & 134.86 &  36.44 &  306.38 & 0.49   & --- \\
28387 &  88 & 169.20 & 54.53 & 208.84 & 121.26 &  481.54 & 0.66   & --- \\
28986 &  66 & 233.43 & 31.07 & 197.06 &  73.09 &  398.63 & 0.69   & \object{A 2092} \\
34726 & 145 & 228.09 &  7.44 & 135.44 & 121.20 &  506.15 & 0.74   & \object{A 2028},\object{A 2033},\object{A 2040} \\
34727 & 256 & 227.74 &  5.78 & 234.02 & 351.58 &  825.94 & 1.25   & \object{A 2028},\object{A 2029},\object{A 2033},\object{A 2040} \\
58305 & 167 & 223.28 & 16.76 & 135.12 & 120.04 &  401.84 & 0.61   & \object{A 1983}\\
62138 & 124 & 223.64 & 18.67 & 175.90 & 127.48 &  456.42 & 0.78   & \object{A 1991} \\
67297 &  95 & 127.24 & 30.48 & 150.20 &  86.98 &  770.39 & 0.45   &  \object{A 671} \\
68625 &  92 & 231.04 & 29.90 & 336.50 & 301.28 &  874.84 & 0.79   & \object{A 2069} \\
73088 & 141 & 215.50 & 48.40 & 213.15 & 184.99 &  631.93 & 0.71   & \object{A 1904} \\
                                                                                            
\label{tab:multicl}                                                                     
\end{tabular}\\                                                                             
\tablefoot{                                                                                   
Columns are as in Table~\ref{tab:unicl}.                                                    
}                                                                                           
\end{table*}                                                                                
                                                                                            
\begin{table*}[ht]
\caption{Results of the tests. Multimodal clusters}
\begin{tabular}{rrrrrcccccccccc} 
\hline\hline  
(1)&(2)&(3)&(4)&(5)& (6)&(7)&(8)&(9)&(10)&(11)&(12)& (13)&(14)&(15) \\      
\hline 
 ID   &$N_{\mathrm{gal}}$ & $v_{\mathrm{pec}}$ & $v_{\mathrm{pec,r}}$ &
$N_{\mathrm{c3D}}$ & $un3D$ & $p_{\mathrm{\Delta}}$ & $p_{\mathrm{\alpha}}$ & $N_{\mathrm{c2D}}$ & $un2D$ &
$p_{\mathrm{\beta}}$ & $p_{\mathrm{AD}}$ & $p_{\mathrm{SW}}$ & $p_{\mathrm{kurt}}$ &  $p_{\mathrm{skew}}$  \\
\hline
  880 &  57 &   400.67 &  0.974 & 3 & $1\cdot10^{-3}$ & 0.002 & 0.002 & 2 & 0.001 & 0.043 & 0.002 & 0.006 & 0.135 & 0.078 \\
  914 & 119 &  -704.39 & -1.071 & 5 & 0.010 & $4\cdot10^{-5}$    & 0.001 & 5 & 0.001 & $1\cdot10^{-3}$ & 0.140 & 0.189 & 0.297 & 0.475 \\
 4122 &  88 & -1091.85 & -1.133 & 3 & 0.002 & $4\cdot10^{-5}$    & $1\cdot10^{-3}$ & 2 & $1\cdot10^{-3}$ & $1\cdot10^{-3}$ & $1\cdot10^{-3}$ & $1\cdot10^{-3}$ & $1\cdot10^{-3}$ & 0.129 \\
11015 &  52 &   140.80 &  0.465 & 8 & $1\cdot10^{-3}$ & 0.037 & 0.232 & 3 & $1\cdot10^{-3}$ & 0.003 & 0.497 & 0.627 & 0.307 & 0.909 \\
11474 &  51 &  -267.37 & -0.873 & 5 & $1\cdot10^{-3}$ & 0.001 & 0.117 & 4 & 0.002 & $1\cdot10^{-3}$ & $1\cdot10^{-3}$ & 0.004 & 0.072 & 0.140 \\
28387 &  88 &  -167.56 & -0.348 & 4 & 0.002 & $4\cdot10^{-5}$    & 0.002 & 3 & 0.001 & $1\cdot10^{-3}$ & 0.001 & 0.017 & 0.217 & 0.469 \\
28986 &  66 &  -205.98 & -0.517 & 5 & 0.003 & 0.002 & 0.002   & 4 & 0.001 & 0.003 & 0.255 & 0.344 & 0.397 & 0.929 \\
34726 & 145 &   -56.63 & -0.112 & 4 & 0.015 & 0.001 & 0.034   & 2 & $1\cdot10^{-3}$ & $1\cdot10^{-3}$ & 0.005 & 0.016 & 0.262 & 0.242 \\
34727 & 256 &  -290.71 & -0.352 & 5 & 0.008 & $4\cdot10^{-5}$ & 0.008    & 6 & 0.017 & $1\cdot10^{-3}$ & 0.042 & 0.071 & 0.120 & 0.337 \\
58305 & 167 &   115.32 &  0.287 & 3 & 0.003 & $4\cdot10^{-5}$ & 0.017    & 4 & 0.009 & $1\cdot10^{-3}$ & $1\cdot10^{-3}$ & $1\cdot10^{-3}$ & 0.010 & 0.007 \\
62138 & 124 &   118.96 &  0.261 & 6 & 0.017 & $4\cdot10^{-5}$ & 0.030    & 4 & 0.046 & 0.392 & 0.238 & 0.189 & 0.912 & 0.754 \\
67297 &  95 &   123.06 &  0.160 & 2 & 0.022 & 0.001 & 0.003   & 2 & 0.106 & 0.003 & 0.002 & 0.001 & 0.356 & 0.116 \\
68625 &  92 &   825.69 &  0.944 & 3 & 0.001 & $4\cdot10^{-5}$ & $1\cdot10^{-3}$  & 2 & $1\cdot10^{-3}$ & 0.040 & 0.003 & 0.015 & 0.017 & 0.427 \\
73088 & 141 &  -224.03 & -0.355 & 3 & 0.014 & $4\cdot10^{-5}$ & $1\cdot10^{-3}$  & 4 & 0.044 & $1\cdot10^{-3}$ & 0.005 & 0.005 & 0.906 & 0.113 \\
\label{tab:multiresults}  
\end{tabular}\\
\tablefoot{                                                                                 
Columns are as in Table~\ref{tab:uniresults}. 
The $p$-value $4\cdot10^{-5}$ denotes the value $p<4\cdot10^{-5}$.
}
\end{table*}
\end{appendix}  

\onltab{1}{
\begin{table*}[ht]
\caption{Data on clusters}
\begin{tabular}{rrrrrrrrr} 
\hline\hline  
(1)&(2)&(3)&(4)&(5)& (6)&(7)&(8)&(9) \\      
\hline 
 ID&$N_{\mathrm{gal}}$& $\mathrm{R.A.}$ & $\mathrm{Dec.}$ &
 $\mathrm{Dist.}$ & $L_{\mathrm{tot}}$ & $\sigma$ & $r_{\mathrm{vir}}$ & Abell ID \\
 & &[deg]&[deg]&[$h^{-1}$ Mpc]&[$10^{10} h^{-2} L_{\sun}$]&$[km s^{-1}$]&[$h^{-1}$ Mpc]& \\
\hline
   18 &  87 & 163.28 & 55.04 & 216.03 & 110.83 &  513.92 & 0.73 & ---   \\
  323 &  67 & 241.45 & 33.34 & 178.18 &  73.10 &  276.83 & 0.67 & ---   \\
  608 &  60 & 245.23 & 29.83 & 284.84 & 132.16 &  532.83 & 0.66 & 2175 \\
  748 &  79 & 159.85 &  5.22 & 206.78 &  93.59 &  748.24 & 0.43 & 1066 \\
  793 & 122 & 234.96 & 21.73 & 123.83 &  63.65 &  515.27 & 0.56 & 2107 \\
  880 &  57 & 211.40 &  6.27 & 249.48 & 101.82 &  411.37 & 0.84 & ---  \\
  914 & 119 & 202.65 & -1.94 & 256.38 & 227.45 &  657.41 & 0.83 & 1750 \\
 1469 &  56 & 151.94 &  0.53 & 287.04 &  99.79 &  418.98 & 0.69 &  933 \\
 1573 &  57 & 153.48 & -0.89 & 136.69 &  35.71 &  744.00 & 0.25 &  ---  \\
 1944 &  60 & 239.76 & 18.06 & 137.41 &  47.99 &  440.72 & 0.46 & ---   \\
 2067 &  62 & 155.48 & 38.53 & 163.84 &  66.62 &  574.01 & 0.44 & ---   \\
 3714 &  82 & 164.58 &  1.56 & 121.07 &  54.71 &  344.93 & 0.60 &  1139 \\
 4122 &  88 & 173.06 & 56.09 & 159.40 &  68.62 &  963.42 & 0.49 &  1291 \\
 4713 &  80 & 176.84 & 55.69 & 155.28 &  80.31 &  637.82 & 0.43 & 1377 \\
 4744 &  71 & 170.78 &  1.05 & 221.36 &  88.70 &  492.96 & 0.62 & 1238 \\
 4992 &  68 & 228.80 &  4.37 & 288.03 & 157.59 &  619.94 & 0.65 &  2048  \\
 5217 &  89 & 180.06 & 56.22 & 193.33 &  94.88 &  577.08 & 0.61 & 1436 \\
 7102 &  54 & 222.21 & 11.27 & 156.93 &  33.32 &  465.99 & 0.46 & ---   \\
 7932 &  50 & 134.55 & 38.49 & 275.41 &  99.94 &  413.99 & 0.55 & 724  \\
 9029 &  78 & 146.58 & 43.18 & 216.96 &  99.94 &  312.14 & 0.72 & 865 \\
 9350 &  89 & 230.41 &  7.71 & 134.14 &  55.05 &  480.48 & 0.65 & 2055, 2063 \\
 9985 &  58 & 152.53 & 54.45 & 138.94 &  41.58 &  384.99 & 0.43 & ---  \\
10438 &  65 & 158.30 & 56.83 & 136.38 &  45.02 &  395.18 & 0.46 & ---   \\
11015 &  52 & 212.51 & 55.07 & 124.69 &  45.80 &  303.10 & 0.46 & ---   \\
11474 &  51 & 218.27 & 52.90 & 134.86 &  36.44 &  306.38 & 0.49 & ---   \\
11683 &  54 & 183.64 & 59.92 & 179.36 &  52.25 &  342.89 & 0.46 & 1507 \\
12508 &  93 & 243.68 & 49.32 & 171.06 &  90.76 &  385.94 & 0.77 & 2169 \\
12540 & 103 & 204.00 & 59.18 & 209.84 & 107.55 &  764.36 & 0.52 & 1767 \\
13216 &  57 & 118.98 & 45.64 & 155.75 &  48.58 &  400.40 & 0.54 & ---   \\
13347 &  50 & 156.40 & 47.59 & 186.67 &  64.31 &  479.41 & 0.48 & 1003 \\
13408 &  58 & 240.25 & 54.01 & 192.66 &  62.67 &  282.97 & 0.73 & 2149  \\
16094 &  71 & 197.79 & 39.26 & 216.08 & 113.02 &  735.05 & 0.50 & 1691 \\
16309 &  69 & 255.66 & 34.03 & 290.56 & 151.63 &  878.96 & 0.46 & 2244 \\
16350 &  65 & 255.67 & 33.50 & 260.44 & 135.10 &  859.11 & 0.36 & 2245   \\
17210 &  72 & 257.43 & 34.48 & 246.34 & 120.38 &  872.61 & 0.47 & 2249 \\
18029 &  53 & 205.69 & 29.94 & 128.59 &  37.05 &  597.71 & 0.41 & 1781 \\
18048 &  78 & 173.28 & 14.42 & 242.21 & 121.96 &  596.51 & 0.61 & ---   \\
20159 &  52 & 158.02 & 40.16 & 203.31 &  59.80 &  517.98 & 0.36 & 1026, 1035 \\
20419 &  58 & 202.31 & 37.53 & 169.15 &  44.07 &  424.97 & 0.59 & 1749 \\
20514 &  56 & 168.82 & 25.83 & 145.55 &  33.03 &  317.82 & 0.42 & ---   \\
21573 &  50 & 176.47 & 15.50 & 205.25 &  60.55 &  364.05 & 0.49 & ---   \\
22572 &  77 & 167.10 & 44.08 & 176.56 &  75.43 &  533.27 & 0.43 & 1169 \\
23374 & 114 & 207.22 & 26.68 & 186.81 & 100.95 &  662.05 & 0.72 & 1795, 1818 \\
23524 &  50 & 207.65 & 29.31 & 228.53 &  63.09 &  304.50 & 0.67 & ---   \\
24554 &  50 & 192.37 & -1.63 & 250.75 &  83.48 &  658.24 & 0.58 & 1620 \\
24604 &  50 & 209.82 & 28.01 & 222.60 &  59.36 &  857.98 & 0.50 & 1831\\
24829 &  77 & 195.63 & -2.57 & 246.69 & 126.42 &  534.45 & 0.66 & 1663\\
25078 &  51 & 194.69 & -1.70 & 249.24 &  87.64 &  498.42 & 0.61 & 1650 \\
28272 &  51 & 248.33 & 11.80 & 153.80 &  40.89 &  355.14 & 0.50 & ---   \\
28387 &  88 & 169.20 & 54.53 & 208.84 & 121.26 &  481.54 & 0.66 & ---   \\
28508 &  58 & 172.42 & 54.10 & 205.02 &  74.36 &  494.72 & 0.40 & 1270 \\
28986 &  66 & 233.43 & 31.07 & 197.06 &  73.09 &  398.63 & 0.69 & 2092 \\
29348 &  75 & 239.19 & 28.65 & 228.83 & 119.40 &  418.12 & 0.69 &  ---  \\
29350 &  55 & 238.13 & 27.69 & 242.59 &  92.61 &  334.90 & 0.64 & ---   \\
29587 & 207 & 239.52 & 27.32 & 264.53 & 365.47 &  740.31 & 0.87 & 2142 \\
29744 &  53 & 219.29 & 24.72 & 261.87 &  96.53 &  396.59 & 0.76 & 1939 \\
30391 &  68 & 248.60 & 26.75 & 209.59 &  93.25 &  271.28 & 0.73 & ---   \\
32006 &  71 & 174.33 & 55.04 & 170.64 &  72.95 &  420.65 & 0.53 & 1396, 1400 \\
32663 &  51 & 225.59 & 21.36 & 186.20 &  61.37 &  435.82 & 0.73 & ---   \\
32909 &  79 & 226.20 & 28.44 & 173.18 &  69.08 &  560.03 & 0.56 & 2022 \\
32976 &  64 & 207.90 &  9.50 & 195.28 &  72.04 &  510.21 & 0.68 & 1808 \\
33082 &  77 & 177.27 & 54.72 & 178.28 &  79.96 &  382.43 & 0.52 & 1383, 1396, 1400 \\
33739 &  79 & 214.39 &  8.24 & 172.84 &  65.40 &  517.22 & 0.49 & 1890 \\
33851 & 138 & 230.43 & 20.75 & 121.03 &  74.65 &  354.61 & 0.77 & ---   \\
34513 &  53 & 225.86 &  7.88 & 261.94 &  94.18 &  426.82 & 0.75 & 2020 \\
\label{tab:cldata1}  
\end{tabular}\\
\tablefoot{                                                                                 
Columns are as follows:
1: ID of the cluster;
2: the number of galaxies in the cluster, $N_{\mathrm{gal}}$;
3--4: cluster right ascension and declination;
5: cluster comoving distance;
6: cluster total luminosity;
7: rms velocity of galaxies in the cluster;
8: cluster virial radius;
9: Abell ID of the cluster.
}
\end{table*}
}

\onltab{1}{
\begin{table*}[ht]
\caption{... continued}
\begin{tabular}{rrrrrrrrr} 
\hline\hline  
(1)&(2)&(3)&(4)&(5)& (6)&(7)&(8)&(9) \\      
\hline 
 ID&$N_{\mathrm{gal}}$& $\mathrm{R.A.}$ & $\mathrm{Dec.}$ 
 &$\mathrm{Dist.}$ & $L_{\mathrm{tot}}$ & $\sigma$ & $r_{\mathrm{vir}}$ & Abell ID \\
 & &[deg]&[deg]&[$h^{-1}$ Mpc]&[$10^{10} h^{-2} L_{\sun}$]&$[km s^{-1}$]&[$h^{-1}$ Mpc]& \\
\hline
34726 & 145 & 228.09 &  7.44 & 135.44 & 121.20 &  506.15 & 0.74   & 2028, 2033, 2040 \\
34727 & 256 & 227.74 &  5.78 & 234.02 & 351.58 &  825.94 & 1.25   & 2028, 2029, 2033, 2040\\
35037 &  79 & 236.16 & 36.16 & 195.74 &  76.68 &  691.85 & 0.58   & 2122, 2124 \\
36861 &  66 & 192.32 & 55.05 & 248.22 & 110.73 &  493.86 & 0.84   & 1616 \\
38087 & 169 & 168.12 & 40.60 & 224.46 & 209.25 &  541.59 & 0.84   & 1173, 1187, 1190, 1203 \\
39489 & 166 & 230.70 & 27.80 & 212.87 & 188.19 & 1061.85 & 0.72   & 2056, 2065\\
39752 & 108 & 231.90 & 28.72 & 195.69 & 116.96 &  514.78 & 0.75   & 2073, 2079 \\
39914 &  63 & 233.12 & 28.02 & 218.38 &  69.77 &  446.50 & 0.65   & 2089  \\
40520 &  52 & 188.89 & 16.52 & 209.16 &  56.00 &  486.67 & 0.67   & 1569 \\
40870 & 118 & 190.34 & 18.56 & 215.43 & 153.64 &  717.36 & 0.69   & ---  \\
42481 &  57 & 246.77 & 14.17 & 151.19 &  55.26 &  354.04 & 0.53   & ---  \\
43336 &  68 & 207.98 & 46.33 & 186.81 &  73.08 &  471.40 & 0.60   & ---  \\
43545 &  51 & 118.25 & 29.41 & 179.41 &  42.75 &  577.85 & 0.51   & 602 \\
43966 &  74 & 196.06 & 19.31 & 190.82 &  76.03 &  613.87 & 0.52   & 1668, 1669 \\
44471 & 113 & 169.12 & 29.26 & 141.53 &  90.86 &  464.08 & 0.55   & 1185, 1213 \\
47492 &  74 & 146.77 & 54.50 & 138.69 &  57.06 &  458.67 & 0.56   & ---  \\
48448 &  55 & 140.25 & 54.85 & 136.34 &  41.93 &  340.67 & 0.54   & 784 \\
50129 &  52 & 205.43 & 26.39 & 225.02 &  61.83 &  449.92 & 0.51   & 1775 \\
50631 & 101 & 116.97 & 18.60 & 142.30 &  86.34 &  636.46 & 0.55   & ---  \\
50647 &  52 & 232.32 & 52.88 & 217.84 &  62.30 &  525.70 & 0.52   & ---  \\
50657 &  55 & 116.50 & 18.24 & 153.75 &  49.96 &  555.58 & 0.42   & ---  \\
52913 &  67 & 129.94 & 28.93 & 237.98 & 116.60 &  368.57 & 0.77   & 690 \\
56571 &  55 & 223.57 & 54.27 & 294.12 & 139.89 &  457.04 & 0.60   & 1999, 2000 \\
57317 & 118 & 216.74 & 16.76 & 158.70 & 105.37 &  516.37 & 0.59   & 1913 \\
58101 & 105 & 168.64 &  2.55 & 228.30 & 122.05 &  614.45 & 0.92   & 1205\\
58305 & 167 & 223.28 & 16.76 & 135.12 & 120.04 &  401.84 & 0.61   & 1983\\
58323 &  64 & 150.66 & 32.72 & 153.13 &  37.74 &  393.28 & 0.40   & --- \\
58604 &  58 & 170.39 &  2.86 & 150.50 &  39.84 &  528.86 & 0.42   & ---  \\
59794 &  90 & 187.45 & 11.71 & 255.93 & 141.55 &  650.79 & 0.75   & 1552 \\
60539 & 107 & 184.40 &  3.65 & 230.10 & 136.74 &  830.47 & 0.55   & 1516  \\
61613 &  77 & 175.58 &  7.85 & 257.83 & 138.13 &  517.01 & 1.06   & 1358 \\
62138 & 124 & 223.64 & 18.67 & 175.90 & 127.48 &  456.42 & 0.78   & 1991 \\
63361 &  72 & 205.59 &  2.26 & 230.57 & 103.06 &  646.10 & 0.64   & 1773 \\
63757 &  87 & 186.99 &  8.79 & 267.64 & 155.17 &  653.32 & 0.65   & 1541 \\
63949 &  80 & 179.25 &  5.06 & 226.17 & 112.03 &  661.43 & 0.55   & 1424 \\
64635 & 109 & 214.18 &  1.95 & 162.93 &  96.38 &  489.43 & 0.73   & ---  \\
64702 &  64 & 184.65 &  5.20 & 227.92 &  84.17 &  539.62 & 0.74   & 1516 \\
67116 &  80 & 208.29 &  5.17 & 236.99 & 114.14 &  651.61 & 0.44   & 1809 \\
67297 &  95 & 127.24 & 30.48 & 150.20 &  86.98 &  770.39 & 0.45   & 671 \\
68376 & 106 & 230.35 & 30.63 & 230.18 & 157.52 &  671.54 & 0.51   & 2061, 2067 \\
68625 &  92 & 231.04 & 29.90 & 336.50 & 301.28 &  874.84 & 0.79   & 2069 \\
73088 & 141 & 215.50 & 48.40 & 213.15 & 184.99 &  631.93 & 0.71   & 1904 \\
73420 &  68 & 122.46 & 35.12 & 247.01 & 105.76 &  555.48 & 0.73   & 628 \\
74783 &  65 & 155.30 & 23.96 & 120.16 &  46.02 &  401.86 & 0.48   & ---  \\

\label{tab:cldata1}  
\end{tabular}\\
\tablefoot{                                                                                 
Columns are as follows:
1: ID of the cluster;
2: the number of galaxies in the cluster, $N_{\mathrm{gal}}$;
3--4: cluster right ascension and declination;
5: cluster comoving distance;
6: cluster total luminosity;
7: rms velocity of galaxies in the cluster;
8: cluster virial radius;
9: Abell ID of the cluster.
}
\end{table*}
}

\onltab{2}{
\begin{table*}[ht]
\caption{Results of the tests}
\begin{tabular}{rrrrrccccccccc} 
\hline\hline  
(1)&(2)&(3)&(4)&(5)& (6)&(7)&(8)&(9)&(10)&(11)&(12)& (13)&(14) \\      
\hline 
 ID    & $v_{\mathrm{pec}}$ & $v_{\mathrm{pec,r}}$ &
$N_{\mathrm{3D}}$ & $un3D$ & $p_{\mathrm{\Delta}}$ & $p_{\mathrm{\alpha}}$ & $N_{\mathrm{2D}}$ & $un2D$ &
$p_{\mathrm{\beta}}$ & $p_{AD}$ & $p_{SW}$ & $p_{kurt}$ &  $p_{skew}$  \\
\hline
   18 &   387.17 &  0.753 & 2 & $1\cdot10^{-3}$ & $4\cdot10^{-5}$ & $1\cdot10^{-3}$ & 2 & $1\cdot10^{-3}$ & 0.006 & 0.229 & 0.429 & 0.856 & 0.638 \\
  323 &   237.15 &  0.857 & 3 & 0.001 & 0.180 & 0.293 & 2 & $1\cdot10^{-3}$ & 0.021 & 0.204 & 0.124 & 0.014 & 0.817 \\
  608 &  -228.43 & -0.429 & 1 & $1\cdot10^{-3}$ & 0.884 & 0.080 & 1 & $1\cdot10^{-3}$ & 0.299 & 0.735 & 0.860 & 0.611 & 0.511 \\
  748 &   613.95 &  0.821 & 1 & $1\cdot10^{-3}$ & 0.007 & 0.007 & 1 & $1\cdot10^{-3}$ & 0.011 & 0.308 & 0.076 & 0.154 & 0.564 \\
  793 &  -384.68 & -0.747 & 3 & 0.019 & $4\cdot10^{-5}$ & 0.008 & 4 & 0.096 & 0.006 & 0.356 & 0.404 & 0.532 & 0.492 \\
  880 &   400.67 &  0.974 & 3 & $1\cdot10^{-3}$ & 0.002 & 0.002 & 2 & 0.001 & 0.043 & 0.002 & 0.006 & 0.135 & 0.078 \\
  914 &  -704.39 & -1.071 & 5 & 0.010 & $4\cdot10^{-5}$ & 0.001 & 5 & 0.001 & $1\cdot10^{-3}$ & 0.140 & 0.189 & 0.297 & 0.475 \\
 1469 &  -342.33 & -0.817 & 2 & 0.005 & 0.198 & 0.101 & 2 & 0.006 & 0.050 & 0.890 & 0.910 & 0.920 & 0.854 \\
 1573 &  1228.28 &  1.651 & 3 & 0.004 & 0.334 & 0.241 & 3 & 0.003 & 0.275 & 0.086 & 0.035 & 0.001 & 0.860 \\
 1944 &  -122.77 & -0.279 & 3 & 0.003 & 0.020 & 0.001 & 2 & $1\cdot10^{-3}$ & $1\cdot10^{-3}$ & 0.104 & 0.026 & 0.349 & 0.114 \\
 2067 &  -788.08 & -1.373 & 2 & $1\cdot10^{-3}$ & 0.007 & 0.016 & 2 & $1\cdot10^{-3}$ & 0.032 & 0.164 & 0.128 & 0.145 & 0.596 \\
 3714 &    -6.37 & -0.018 & 2 & 0.013 & $4\cdot10^{-5}$ & $1\cdot10^{-3}$ & 1 & $1\cdot10^{-3}$ & 0.057 & 0.816 & 0.552 & 0.992 & 0.478 \\
 4122 & -1091.85 & -1.133 & 3 & 0.002 & $4\cdot10^{-5}$ & $1\cdot10^{-3}$ & 2 & $1\cdot10^{-3}$ & $1\cdot10^{-3}$ & $1\cdot10^{-3}$ & $1\cdot10^{-3}$ & $1\cdot10^{-3}$ & 0.129 \\
 4713 &   -78.38 & -0.123 & 3 & 0.016 & 0.003 & $1\cdot10^{-3}$ & 2 & $1\cdot10^{-3}$ & 0.003 & 0.586 & 0.766 & 0.840 & 0.825 \\
 4744 &   147.05 &  0.298 & 4 & $1\cdot10^{-3}$ & 0.001 & 0.002 & 3 & 0.012 & 0.015 & 0.954 & 0.973 & 0.828 & 0.994 \\
 4992 &   303.46 &  0.489 & 2 & $1\cdot10^{-3}$ & 0.048 & 0.007 & 2 & $1\cdot10^{-3}$ & 0.934 & 0.863 & 0.700 & 0.655 & 0.637 \\
 5217 &   -19.11 & -0.033 & 1 & $1\cdot10^{-3}$ & 0.216 & 0.110 & 1 & $1\cdot10^{-3}$ & 0.367 & 0.240 & 0.309 & 0.136 & 0.658 \\
 7102 &   326.16 &  0.700 & 2 & $1\cdot10^{-3}$ & 0.022 & 0.658 & 1 & $1\cdot10^{-3}$ & 0.097 & 0.162 & 0.329 & 0.407 & 0.421 \\
 7932 &  -111.73 & -0.270 & 2 & $1\cdot10^{-3}$ & 0.089 & 0.076 & 2 & $1\cdot10^{-3}$ & 0.007 & 0.985 & 0.998 & 0.421 & 0.797 \\
 9029 &  -341.66 & -1.095 & 2 & 0.005 & 0.094 & 0.069 & 2 & 0.004 & 0.616 & 0.697 & 0.696 & 0.262 & 0.909 \\
 9350 &  -129.46 & -0.269 & 4 & $1\cdot10^{-3}$ & 0.020 & 0.155 & 1 & $1\cdot10^{-3}$ & 0.068 & 0.392 & 0.373 & 0.718 & 0.470 \\
 9985 &    57.85 &  0.150 & 2 & 0.014 & 0.167 & 0.414 & 2 & 0.002 & 0.953 & 0.252 & 0.339 & 0.251 & 0.285 \\
10438 &  -161.54 & -0.409 & 2 & 0.001 & 0.008 & 0.036 & 2 & $1\cdot10^{-3}$ & 0.003 & 0.154 & 0.096 & 0.008 & 0.704 \\
11015 &   140.80 &  0.465 & 8 & $1\cdot10^{-3}$ & 0.037 & 0.232 & 3 & $1\cdot10^{-3}$ & 0.003 & 0.497 & 0.627 & 0.307 & 0.909 \\
11474 &  -267.37 & -0.873 & 5 & $1\cdot10^{-3}$ & 0.001 & 0.117 & 4 & 0.002 & $1\cdot10^{-3}$ & $1\cdot10^{-3}$ & 0.004 & 0.072 & 0.140 \\
11683 &   -21.81 & -0.064 & 1 & $1\cdot10^{-3}$ & 0.064 & 0.027 & 1 & $1\cdot10^{-3}$ & 0.296 & 0.814 & 0.862 & 0.826 & 0.874 \\
12508 &   -78.19 & -0.203 & 2 & $1\cdot10^{-3}$ & $4\cdot10^{-5}$ & 0.016 & 3 & 0.030 & $1\cdot10^{-3}$ & 0.929 & 0.962 & 0.447 & 0.893 \\
12540 &   477.21 &  0.624 & 2 & 0.037 & 0.345 & 0.024 & 2 & 0.082 & 0.015 & 0.006 & 0.023 & 0.713 & 0.197 \\
13216 &   261.75 &  0.654 & 3 & $1\cdot10^{-3}$ & $4\cdot10^{-5}$ & $1\cdot10^{-3}$ & 3 & $1\cdot10^{-3}$ & $1\cdot10^{-3}$ & 0.126 & 0.109 & 0.146 & 0.538 \\
13347 &   369.53 &  0.771 & 2 & $1\cdot10^{-3}$ & 0.007 & 0.008 & 2 & $1\cdot10^{-3}$ & $1\cdot10^{-3}$ & 0.159 & 0.323 & 0.219 & 0.539 \\
13408 &   114.55 &  0.405 & 2 & $1\cdot10^{-3}$ & 0.752 & 0.782 & 2 & 0.004 & 0.002 & 0.312 & 0.311 & 0.283 & 0.236 \\
16094 &   -47.45 & -0.065 & 3 & 0.001 & 0.021 & 0.117 & 2 & 0.003 & 0.292 & 0.687 & 0.674 & 0.739 & 0.562 \\
16309 &    84.78 &  0.096 & 4 & 0.023 & 0.004 & 0.007 & 2 & $1\cdot10^{-3}$ & 0.022 & 0.058 & 0.034 & 0.171 & 0.308 \\
16350 &  -626.62 & -0.729 & 2 & $1\cdot10^{-3}$ & $4\cdot10^{-5}$ & 0.004 & 2 & 0.008 & 0.429 & 0.011 & 0.017 & $1\cdot10^{-3}$ & 0.900 \\
17210 & -1308.10 & -1.499 & 2 & $1\cdot10^{-3}$ & 0.007 & 0.017 & 1 & $1\cdot10^{-3}$ & 0.041 & 0.162 & 0.214 & 0.780 & 0.997 \\
18029 &  -625.44 & -1.046 & 3 & $1\cdot10^{-3}$ & $4\cdot10^{-5}$ & $1\cdot10^{-3}$ & 2 & $1\cdot10^{-3}$ & 0.015 & 0.155 & 0.182 & 0.069 & 0.649 \\
18048 &    28.39 &  0.048 & 2 & $1\cdot10^{-3}$ & 0.010 & 0.003 & 3 & $1\cdot10^{-3}$ & $1\cdot10^{-3}$ & 0.019 & 0.083 & 0.200 & 0.417 \\
20159 &  -300.41 & -0.580 & 2 & 0.011 & 0.011 & 0.005 & 1 & $1\cdot10^{-3}$ & 0.875 & 0.430 & 0.224 & 0.087 & 0.657 \\
20419 &   201.67 &  0.475 & 3 & 0.001 & 0.004 & 0.019 & 1 & $1\cdot10^{-3}$ & 0.114 & 0.011 & 0.042 & 0.573 & 0.252 \\
20514 &  -390.07 & -1.227 & 4 & $1\cdot10^{-3}$ & 0.436 & 0.041 & 3 & $1\cdot10^{-3}$ & 0.003 & 0.753 & 0.816 & 0.536 & 0.907 \\
21573 &   -32.47 & -0.089 & 2 & 0.001 & 0.488 & 0.059 & 2 & $1\cdot10^{-3}$ & 0.006 & 0.496 & 0.410 & 0.648 & 0.583 \\
22572 &   -33.79 & -0.063 & 2 & $1\cdot10^{-3}$ & $4\cdot10^{-5}$ & $1\cdot10^{-3}$ & 2 & $1\cdot10^{-3}$ & 0.042 & 0.064 & 0.035 & 0.016 & 0.656 \\
23374 &  1226.48 &  1.853 & 2 & $1\cdot10^{-3}$ & 0.001 & 0.052 & 2 & $1\cdot10^{-3}$ & 0.135 & 0.038 & 0.026 & 0.005 & 0.716 \\
23524 &   -68.68 & -0.226 & 2 & $1\cdot10^{-3}$ & 0.166 & 0.213 & 2 & $1\cdot10^{-3}$ & 0.027 & 0.023 & 0.009 & 0.047 & 0.071 \\
24554 &   107.53 &  0.163 & 3 & 0.001 & $4\cdot10^{-5}$ & $1\cdot10^{-3}$ & 2 & $1\cdot10^{-3}$ & 0.021 & 0.183 & 0.133 & 0.054 & 0.737 \\
24604 &  -256.27 & -0.299 & 2 & $1\cdot10^{-3}$ & 0.001 & 0.008 & 1 & $1\cdot10^{-3}$ & 0.662 & 0.079 & 0.049 & 0.021 & 0.562 \\
24829 &  -107.41 & -0.201 & 3 & 0.004 & 0.003 & $1\cdot10^{-3}$ & 2 & 0.016 & 0.091 & 0.240 & 0.201 & 0.592 & 0.496 \\
25078 &  -120.45 & -0.242 & 1 & $1\cdot10^{-3}$ & 0.270 & 0.378 & 1 & $1\cdot10^{-3}$ & 0.829 & 0.258 & 0.513 & 0.753 & 0.452 \\
28272 &   464.63 &  1.308 & 1 & $1\cdot10^{-3}$ & 0.032 & 0.213 & 3 & 0.004 & 0.645 & 0.244 & 0.307 & 0.568 & 0.298 \\
28387 &  -167.56 & -0.348 & 4 & 0.002 & $4\cdot10^{-5}$ & 0.002 & 3 & 0.001 & $1\cdot10^{-3}$ & 0.001 & 0.017 & 0.217 & 0.469 \\
28508 &   478.41 &  0.967 & 1 & $1\cdot10^{-3}$ & 0.482 & 0.073 & 1 & $1\cdot10^{-3}$ & 0.272 & 0.004 & 0.004 & 0.992 & 0.114 \\
28986 &  -205.98 & -0.517 & 5 & 0.003 & 0.002 & 0.002 & 4 & 0.001 & 0.003 & 0.255 & 0.344 & 0.397 & 0.929 \\
29348 &     9.90 &  0.024 & 4 & 0.002 & 0.001 & 0.067 & 4 & 0.006 & $1\cdot10^{-3}$ & 0.910 & 0.962 & 0.550 & 0.806 \\
29350 &  -309.94 & -0.925 & 2 & 0.003 & 0.325 & 0.956 & 3 & 0.007 & 0.168 & 0.964 & 0.897 & 0.760 & 0.988 \\
29587 &   334.44 &  0.452 & 3 & 0.008 & $4\cdot10^{-5}$ & $1\cdot10^{-3}$ & 3 & 0.140 & $1\cdot10^{-3}$ & 0.178 & 0.170 & 0.687 & 0.271 \\
29744 &  -557.37 & -1.405 & 3 & 0.003 & 0.004 & 0.253 & 2 & $1\cdot10^{-3}$ & 0.045 & 0.628 & 0.787 & 0.291 & 0.850 \\
30391 &   279.63 &  1.031 & 3 & $1\cdot10^{-3}$ & $4\cdot10^{-5}$ & 0.003 & 3 & $1\cdot10^{-3}$ & 0.005 & 0.165 & 0.312 & 0.395 & 0.284 \\
32006 &    11.99 &  0.029 & 2 & $1\cdot10^{-3}$ & $4\cdot10^{-5}$ & 0.001 & 3 & $1\cdot10^{-3}$ & $1\cdot10^{-3}$ & 0.914 & 0.824 & 0.476 & 0.851 \\
32663 &    96.03 &  0.220 & 1 & $1\cdot10^{-3}$ & 0.008 & 0.147 & 1 & $1\cdot10^{-3}$ & 0.564 & 0.060 & 0.041 & 0.174 & 0.118 \\
32909 &  -223.60 & -0.399 & 3 & 0.025 & 0.006 & 0.009 & 2 & 0.005 & 0.006 & 0.910 & 0.895 & 0.866 & 0.700 \\
32976 &   157.29 &  0.308 & 2 & 0.016 & 0.053 & $1\cdot10^{-3}$ & 1 & $1\cdot10^{-3}$ & 0.916 & 0.258 & 0.224 & 0.640 & 0.430 \\
33082 &  -137.51 & -0.360 & 2 & $1\cdot10^{-3}$ & 0.120 & 0.081 & 4 & $1\cdot10^{-3}$ & $1\cdot10^{-3}$ & 0.456 & 0.596 & 0.912 & 0.787 \\
33739 &  -217.81 & -0.421 & 4 & 0.070 & 0.003 & 0.035 & 2 & 0.002 & 0.024 & 0.054 & 0.136 & 0.440 & 0.451 \\
33851 &   472.01 &  1.331 & 3 & 0.002 & 0.057 & 0.496 & 3 & 0.002 & 0.037 & 0.153 & 0.200 & 0.768 & 0.221 \\
34513 &   176.44 &  0.413 & 2 & $1\cdot10^{-3}$ & 0.004 & 0.008 & 1 & $1\cdot10^{-3}$ & 0.514 & 0.584 & 0.830 & 0.199 & 0.561 \\
\label{tab:testresults1}  
\end{tabular}\\
\tablefoot{                                                                                 
Columns are as follows:
1: ID of a cluster;
2: peculiar velocity of the main galaxy ($km s^{-1}$),
3: normalised peculiar velocity of the main galaxy,
$v_{\mathrm{pec}}$/$\sigma_{\mathrm{v}}$;
4: the number of components in 3D in a cluster, $N_{\mathrm{comp}}$;
5: the median uncertainty of classification;
6: $p$-value of DS test;
7: $p$-value for $\alpha$ test;
8: the number of components in 2D in a cluster, $N_{\mathrm{comp}}$;
9: the median uncertainty of classification in 2D;
10: $p$-value of $\beta$ test;
11: $p$-value for AD test;
12: $p$-value of SW test;
13: $p$-value for $kurtosis$ test;
14: $p$-value for $skewness$ test.
}
\end{table*}
}

\onltab{2}{
\begin{table*}[ht]
\caption{... continued}
\begin{tabular}{rrrrrccccccccc} 
\hline\hline  
(1)&(2)&(3)&(4)&(5)& (6)&(7)&(8)&(9)&(10)&(11)&(12)& (13)&(14) \\      
\hline 
 ID    & $v_{\mathrm{pec}}$ & $v_{\mathrm{pec,r}}$ &
$N_{\mathrm{3D}}$ & $un3D$ & $p_{\mathrm{\Delta}}$ & $p_{\mathrm{\alpha}}$ & $N_{\mathrm{2D}}$ & $un2D$ &
$p_{\mathrm{\beta}}$ & $p_{AD}$ & $p_{SW}$ & $p_{kurt}$ &  $p_{skew}$  \\
\hline
34726 &   -56.63 & -0.112 & 4 & 0.015 & 0.001 & 0.034 & 2 & $1\cdot10^{-3}$ & $1\cdot10^{-3}$ & 0.005 & 0.016 & 0.262 & 0.242 \\
34727 &  -290.71 & -0.352 & 5 & 0.008 & $4\cdot10^{-5}$ & 0.008 & 6 & 0.017 & $1\cdot10^{-3}$ & 0.042 & 0.071 & 0.120 & 0.337 \\
35037 &   748.65 &  1.082 & 4 & 0.002 & 0.028 & 0.002 & 2 & 0.003 & 0.002 & 0.163 & 0.047 & 0.646 & 0.178 \\
36861 &   379.45 &  0.768 & 3 & 0.001 & $4\cdot10^{-5}$ & 0.004 & 2 & $1\cdot10^{-3}$ & $1\cdot10^{-3}$ & 0.213 & 0.255 & 0.483 & 0.304 \\
38087 &   844.40 &  1.559 & 2 & $1\cdot10^{-3}$ & 0.031 & 0.003 & 4 & 0.018 & $1\cdot10^{-3}$ & 0.524 & 0.724 & 0.823 & 0.491 \\
39489 &  -521.64 & -0.491 & 3 & 0.046 & $4\cdot10^{-5}$ & $1\cdot10^{-3}$ & 3 & $1\cdot10^{-3}$ & $1\cdot10^{-3}$ & 0.212 & 0.116 & 0.309 & 0.426 \\
39752 &   146.15 &  0.284 & 2 & $1\cdot10^{-3}$ & 0.004 & 0.018 & 2 & 0.101 & 0.013 & 0.207 & 0.351 & 0.690 & 0.770 \\
39914 &   148.99 &  0.334 & 1 & $1\cdot10^{-3}$ & 0.051 & 0.106 & 1 & $1\cdot10^{-3}$ & 0.804 & 0.904 & 0.863 & 0.566 & 0.759 \\
40520 &  -348.16 & -0.715 & 2 & $1\cdot10^{-3}$ & $4\cdot10^{-5}$ & $1\cdot10^{-3}$ & 2 & $1\cdot10^{-3}$ & 0.050 & 0.518 & 0.271 & 0.206 & 0.580 \\
40870 &  -304.93 & -0.425 & 2 & $1\cdot10^{-3}$ & 0.001 & 0.006 & 4 & 0.043 & 0.008 & 0.344 & 0.538 & 0.695 & 0.548 \\
42481 &  -184.87 & -0.522 & 3 & 0.004 & $4\cdot10^{-5}$ & $1\cdot10^{-3}$ & 3 & 0.009 & 0.749 & 0.936 & 0.677 & 0.467 & 0.843 \\
43336 &   -97.43 & -0.207 & 1 & $1\cdot10^{-3}$ & 0.003 & $1\cdot10^{-3}$ & 1 & $1\cdot10^{-3}$ & 0.335 & 0.980 & 0.941 & 0.756 & 0.782 \\
43545 &    81.55 &  0.141 & 1 & $1\cdot10^{-3}$ & 0.004 & 0.011 & 1 & $1\cdot10^{-3}$ & 0.055 & 0.047 & 0.200 & 0.817 & 0.369 \\
43966 &   -10.04 & -0.016 & 2 & $1\cdot10^{-3}$ & 0.155 & 0.015 & 2 & 0.038 & 0.009 & 0.992 & 0.994 & 0.845 & 0.757 \\
44471 &  -411.48 & -0.887 & 4 & 0.002 & $4\cdot10^{-5}$ & 0.025 & 4 & 0.003 & 0.021 & 0.194 & 0.422 & 0.299 & 0.742 \\
47492 &   184.86 &  0.403 & 3 & 0.004 & 0.307 & 0.289 & 3 & 0.034 & 0.179 & 0.723 & 0.884 & 0.938 & 0.731 \\
48448 &   192.05 &  0.564 & 2 & $1\cdot10^{-3}$ & 0.085 & 0.195 & 2 & 0.033 & 0.666 & 0.548 & 0.547 & 0.930 & 0.421 \\
50129 &  -129.26 & -0.287 & 1 & $1\cdot10^{-3}$ & 0.326 & 0.737 & 1 & $1\cdot10^{-3}$ & 0.357 & 0.446 & 0.421 & 0.181 & 0.770 \\
50631 &  -144.67 & -0.227 & 2 & 0.001 & 0.032 & 0.054 & 2 & 0.002 & 0.004 & $1\cdot10^{-3}$ & 0.003 & 0.200 & 0.219 \\
50647 &  -195.09 & -0.371 & 2 & $1\cdot10^{-3}$ & 0.259 & 0.318 & 1 & $1\cdot10^{-3}$ & 0.520 & 0.425 & 0.321 & 0.858 & 0.361 \\
50657 &   427.95 &  0.770 & 1 & $1\cdot10^{-3}$ & 0.078 & 0.401 & 2 & $1\cdot10^{-3}$ & 0.126 & 0.001 & 0.007 & 0.278 & 0.148 \\
52913 &  -287.29 & -0.779 & 3 & $1\cdot10^{-3}$ & 0.011 & 0.050 & 3 & $1\cdot10^{-3}$ & $1\cdot10^{-3}$ & 0.085 & 0.072 & 0.953 & 0.187 \\
56571 &  -368.74 & -0.807 & 4 & $1\cdot10^{-3}$ & 0.016 & 0.037 & 1 & $1\cdot10^{-3}$ & 0.322 & 0.009 & 0.021 & 0.772 & 0.169 \\
57317 &  -712.88 & -1.381 & 2 & 0.001 & $4\cdot10^{-5}$ & 0.007 & 2 & $1\cdot10^{-3}$ & 0.003 & 0.056 & 0.049 & 0.001 & 0.900 \\
58101 &    26.98 &  0.044 & 4 & 0.001 & $4\cdot10^{-5}$ & $1\cdot10^{-3}$ & 3 & $1\cdot10^{-3}$ & $1\cdot10^{-3}$ & 0.398 & 0.257 & 0.162 & 0.858 \\
58305 &   115.32 &  0.287 & 3 & 0.003 & $4\cdot10^{-5}$ & 0.017 & 4 & 0.009 & $1\cdot10^{-3}$ & $1\cdot10^{-3}$ & $1\cdot10^{-3}$ & 0.010 & 0.007 \\
58323 &   -56.43 & -0.143 & 3 & $1\cdot10^{-3}$ & 0.259 & 0.263 & 3 & 0.001 & 0.315 & 0.842 & 0.484 & 0.368 & 0.800 \\
58604 &  -201.00 & -0.380 & 1 & $1\cdot10^{-3}$ & 0.504 & 0.232 & 1 & $1\cdot10^{-3}$ & 0.222 & 0.522 & 0.677 & 0.363 & 0.847 \\
59794 &  -881.78 & -1.355 & 2 & $1\cdot10^{-3}$ & $4\cdot10^{-5}$ & 0.002 & 1 & $1\cdot10^{-3}$ & 0.006 & 0.841 & 0.911 & 0.931 & 0.613 \\
60539 &   -24.93 & -0.030 & 1 & $1\cdot10^{-3}$ & 0.021 & 0.002 & 1 & $1\cdot10^{-3}$ & 0.563 & 0.776 & 0.645 & 0.882 & 0.479 \\
61613 &   100.00 &  0.193 & 4 & 0.001 & $4\cdot10^{-5}$ & 0.001 & 3 & $1\cdot10^{-3}$ & 0.561 & 0.013 & 0.043 & 0.294 & 0.304 \\
62138 &   118.96 &  0.261 & 6 & 0.017 & $4\cdot10^{-5}$ & 0.030 & 4 & 0.046 & 0.392 & 0.238 & 0.189 & 0.912 & 0.754 \\
63361 &  -175.06 & -0.271 & 1 & $1\cdot10^{-3}$ & 0.004 & 0.179 & 1 & $1\cdot10^{-3}$ & 0.278 & 0.240 & 0.455 & 0.347 & 0.820 \\
63757 &  -159.98 & -0.245 & 4 & $1\cdot10^{-3}$ & $4\cdot10^{-5}$ & $1\cdot10^{-3}$ & 3 & $1\cdot10^{-3}$ & $1\cdot10^{-3}$ & 0.056 & 0.070 & 0.047 & 0.429 \\
63949 &  -843.52 & -1.275 & 3 & 0.003 & 0.001 & 0.001 & 2 & 0.001 & 0.222 & 0.332 & 0.203 & 0.218 & 0.414 \\
64635 &  -347.86 & -0.711 & 3 & 0.016 & 0.001 & 0.009 & 3 & 0.006 & $1\cdot10^{-3}$ & 0.145 & 0.105 & 0.150 & 0.132 \\
64702 &   575.72 &  1.067 & 2 & $1\cdot10^{-3}$ & 0.001 & 0.008 & 2 & $1\cdot10^{-3}$ & 0.562 & $1\cdot10^{-3}$ & $1\cdot10^{-3}$ & 0.544 & 0.056 \\
67116 &  -183.37 & -0.281 & 1 & $1\cdot10^{-3}$ & 0.094 & 0.089 & 2 & 0.149 & 0.342 & 0.929 & 0.805 & 0.930 & 0.862 \\
67297 &   123.06 &  0.160 & 2 & 0.022 & 0.001 & 0.003 & 2 & 0.106 & 0.003 & 0.002 & 0.001 & 0.356 & 0.116 \\
68376 &   327.64 &  0.488 & 3 & 0.053 & 0.205 & 0.066 & 1 & $1\cdot10^{-3}$ & 0.146 & $1\cdot10^{-3}$ & 0.001 & 0.848 & 0.061 \\
68625 &   825.69 &  0.944 & 3 & 0.001 & $4\cdot10^{-5}$ & $1\cdot10^{-3}$ & 2 & $1\cdot10^{-3}$ & 0.040 & 0.003 & 0.015 & 0.017 & 0.427 \\
73088 &  -224.03 & -0.355 & 3 & 0.014 & $4\cdot10^{-5}$ & $1\cdot10^{-3}$ & 4 & 0.044 & $1\cdot10^{-3}$ & 0.005 & 0.005 & 0.906 & 0.113 \\
73420 &   221.40 &  0.399 & 3 & 0.003 & $4\cdot10^{-5}$ & $1\cdot10^{-3}$ & 3 & 0.156 & 0.586 & 0.521 & 0.366 & 0.335 & 0.647 \\
74783 &   -55.95 & -0.139 & 3 & 0.004 & 0.138 & 0.007 & 3 & 0.004 & 0.010 & 0.809 & 0.786 & 0.520 & 0.564 \\
\label{tab:testresults1}  
\end{tabular}\\
\tablefoot{                                                                                 
Columns are as follows:
1: ID of a cluster;
2: peculiar velocity of the main galaxy ($km s^{-1}$),
3: normalised peculiar velocity of the main galaxy,
$v_{\mathrm{pec}}$/$\sigma_{\mathrm{v}}$;
4: the number of components in 3D in a cluster, $N_{\mathrm{comp}}$;
5: the median uncertainty of classification;
6: $p$-value of DS test;
7: $p$-value for $\alpha$ test;
8: the number of components in 2D in a cluster, $N_{\mathrm{comp}}$;
9: the median uncertainty of classification in 2D;
10: $p$-value of $\beta$ test;
11: $p$-value for AD test;
12: $p$-value of SW test;
13: $p$-value for $kurtosis$ test;
14: $p$-value for $skewness$ test.
}
\end{table*}
}
\end{document}